\documentclass[letterpaper]{article} 
\usepackage[draft]{aaai2026}  
\usepackage{times}  
\usepackage{helvet}  
\usepackage{courier}  
\usepackage[hyphens]{url}  
\usepackage{graphicx} 

\usepackage{listings}
\usepackage{xcolor}
\usepackage{multirow}
\usepackage{booktabs}
\usepackage{graphicx}
\usepackage{amsmath} 
\usepackage{booktabs}
\usepackage{amssymb}
\usepackage[switch]{lineno}
\usepackage{multirow}
\usepackage{array}
\usepackage{bbding}
\usepackage{longtable}

\usepackage{natbib}  
\usepackage{caption} 
\usepackage{algorithm}
\usepackage{algorithmic}

\usepackage{newfloat}
\usepackage{listings}
\DeclareCaptionStyle{ruled}{labelfont=normalfont,labelsep=colon,strut=off} 
\floatstyle{ruled}
\newfloat{listing}{tb}{lst}{}
\floatname{listing}{Listing}
\title{ComfyGPT: A Self-Optimizing Multi-Agent System for Comprehensive ComfyUI Workflow Generation}

\author{
    Oucheng Huang\textsuperscript{\rm 1\equalcontrib},
    Yuhang Ma\textsuperscript{\rm 2\equalcontrib \thanks{Project Lead}},
    Zeng Zhao\textsuperscript{\rm 2\thanks{Equal advising}},
    Mingrui Wu\textsuperscript{\rm 1},\\
    Jiayi Ji\textsuperscript{\rm 1},
    Rongsheng Zhang\textsuperscript{\rm 2},
    Zhipeng Hu\textsuperscript{\rm 2},
    Xiaoshuai Sun\textsuperscript{\rm 1\footnotemark[3]},
    Rongrong Ji\textsuperscript{\rm 1}
}
\affiliations{
    \textsuperscript{\rm 1}Key Laboratory of Multimedia Trusted Perception and Efficient Computing, \\
    Ministry of Education of China, Xiamen University\\
    \textsuperscript{\rm 2}Fuxi AI Lab, NetEase Inc.\\
}

\usepackage{bibentry}

\begin{document}

\maketitle

\begin{abstract}
ComfyUI is a popular workflow-based interface that allows users to customize image generation tasks through an intuitive node-based system. However, the complexity of managing node connections and diverse modules can be challenging for users. In this paper, we introduce \textbf{ComfyGPT}, a self-optimizing multi-agent system designed to generate ComfyUI workflows based on task descriptions automatically. 
The key innovations of ComfyGPT include: (1) consisting of four specialized agents to build a multi-agent workflow generation system: ReformatAgent, FlowAgent, RefineAgent, and ExecuteAgent; (2) focusing on generating precise node connections instead of entire workflows, improving generation accuracy; and (3) enhancing workflow generation through reinforcement learning. Moreover, we introduce FlowDataset, a large-scale dataset containing $13,571$ workflow-description pairs, and FlowBench, a comprehensive benchmark for evaluating workflow generation systems. Additionally, we propose four novel evaluation metrics: Format Validation (FV), Pass Accuracy (PA), Pass Instruct Alignment (PIA), and Pass Node Diversity (PND). Experimental results demonstrate that ComfyGPT significantly outperforms existing LLM-based methods in workflow generation, making it a significant step forward in this field. Code is avaliable at https://github.com/comfygpt/comfygpt.
\end{abstract}    
\section{Introduction}

In recent years, computer vision drives advances in diverse applications such as text-to-image generation~\cite{StableDiffusion,DALL-E2,Imagen,Infinity,liu2024llm4genleveragingsemanticrepresentation}, image editing~\cite{ip-adapter,xia2023diffir,sabini2018painting,deng2018arcface,characteradapter,Storynizor}, and video generation~\cite{cogvideo,animatediff}.
However, achieving high-quality results often requires the seamless integration of multiple modules. For example, virtual try-on tasks may involve combining text-to-image generation, segmentation models~\cite{kirillov2023segment}, and inpainting models into a single pipeline.

Traditionally, Python scripts are used to connect these components, but managing the complexity of integrating specialized tools poses significant challenges. Developers face technical difficulties in coordinating the system, while end-users often struggle to find intuitive and accessible tools. To address this, community-led solutions like WebUI platforms have emerged, offering visualization-based interfaces that simplify workflows and make them more user-friendly. However, these platforms usually lack the advanced customization needed for complex tasks and cannot fully automate end-to-end generation workflows.

\begin{figure}[t]
  \centering
  \includegraphics[width=1\linewidth]{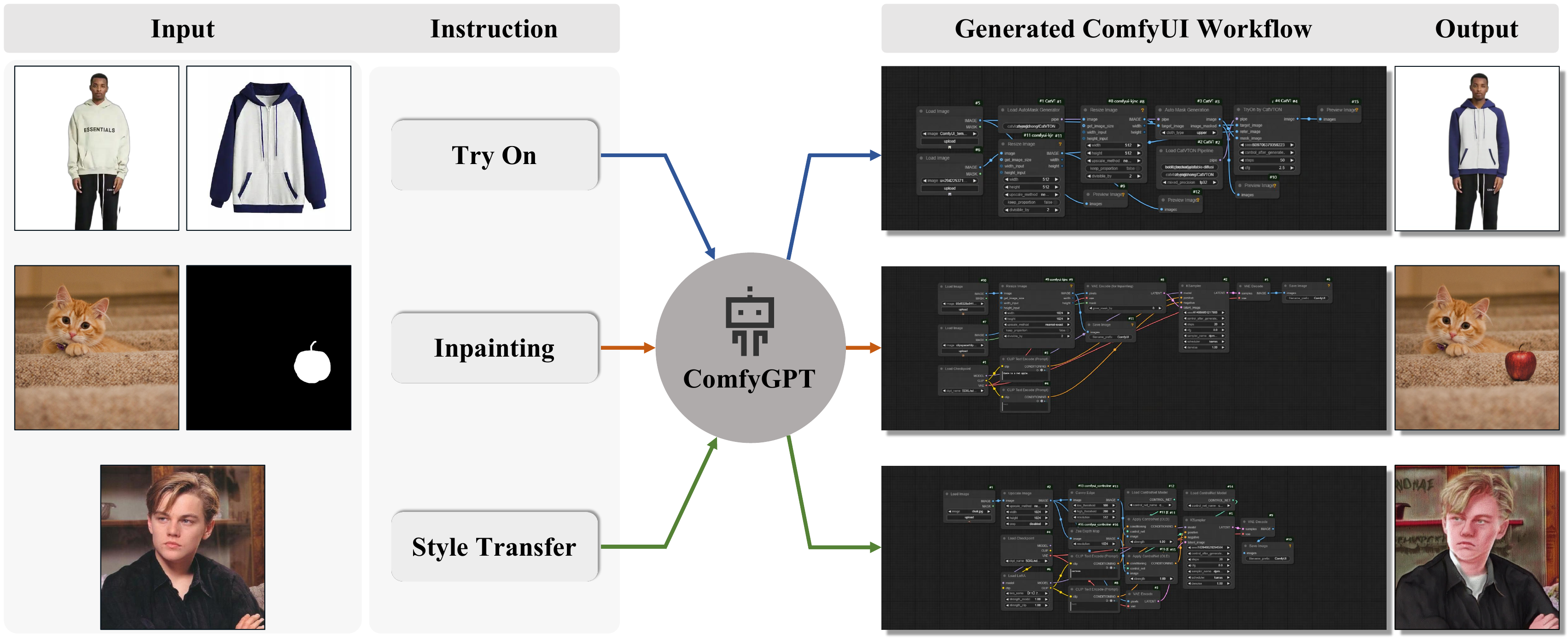}
\caption{\textbf{Workflow generated by ComfyGPT across various task instructions.} By leveraging its strong alignment capabilities, ComfyGPT enables users to generate diverse workflows in response to task instructions.}
  \label{fig:intro}
  \vspace{-4mm}
\end{figure}

ComfyUI offers a versatile framework for workflow-based design, providing modular connectivity through independent nodes without requiring extensive coding. By treating each functional component as a distinct node, it forms flexible topological workflows through connections between these nodes, enabling various functionalities. This structure intuitively reflects the relationships between components and allows users to easily modify workflows by adding, deleting, or replacing modules. However, the complexity of node connections and the diversity of available modules often make it challenging for users to navigate. Additionally, while existing workflows can be reused for similar projects, adapting them to different tasks often requires substantial modifications, limiting their overall transferability.

Inspired by the remarkable capabilities of large language models (LLM)~\cite{language,palm,training,opt,glm,llama,wei2022emergent,deepseekr1},
we introduce \textbf{ComfyGPT}, a self-optimizing multi-agent system designed to streamline workflow creation for ComfyUI through automation. It automatically generates modular workflows based on user instructions, enabling a broad range of tasks.

ComfyGPT introduces two key innovations. First, it employs an advanced self-optimizing multi-agent framework, leveraging LLMs with GRPO~\cite{deepseekr1}. This setup facilitates autonomous error correction and iterative improvement, allowing it to continually refine workflow generation through closed-loop learning. Second, rather than focusing on entire workflows, ComfyGPT breaks the generation process into smaller subtasks, concentrating on individual link connections between nodes. This modular approach enhances the model's capability to handle complex interdependencies and replicate intricate computational graphs more effectively.
\begin{table}[t]
\centering
\resizebox{\columnwidth}{!}{
\begin{tabular}{c|c|cccccc|ccc}
\hline
\multirow{2}{*}{\textbf{Dataset}} & \multirow{2}{*}{\textbf{Approach}} & \multicolumn{6}{c|}{\textbf{Task Categories}} & \multicolumn{3}{c}{\textbf{Dataset Scale}} \\\cmidrule(lr){3-8}\cmidrule(lr){9-11}
 &  & \textbf{T2I} & \textbf{IE} & \textbf{ST} & \textbf{3DG} & \textbf{VGE} & \textbf{O} & \textbf{Instructions} & \textbf{Workflows} & \textbf{Node Types} \\
\hline
\multirow{3}{*}{Training Set}&ComfyGen~\cite{comfygen}&\checkmark& $\times$& $\times$&$\times$ &$\times$ & $\times$&500& 310& -\\
&\textbf{FlowDataset(Ours)}&\checkmark& \checkmark& \checkmark& \checkmark& \checkmark& \checkmark&\textbf{12,571}& \textbf{12,571}& \textbf{3,526}\\
\midrule
\multirow{3}{*} {Benchmark}
&ComfyGen~\cite{comfygen}&\checkmark& $\times$& $\times$&$\times$&$\times$ & $\times$& 500& 0&0\\

&ComfyBench~\cite{comfybench}& \checkmark& \checkmark& \checkmark& $\times$& \checkmark& $\times$&200& 20& 63\\
&\textbf{FlowBench(Ours)} &\checkmark&\checkmark& \checkmark& \checkmark&\checkmark& \checkmark&\textbf{1,000}&\textbf{1,000}&\textbf{745}\\
\hline
\end{tabular}}

\caption{\textbf{Comparison of task categories between our proposed dataset and others in text-to-image generation (\textit{T2I}), image editing (\textit{IE}), style transfer (\textit{ST}), 3D generation (\textit{3DG}), video generation (\textit{VG}), and others (\textit{O}).} 
The Dataset Scale column summarizes the number of instructions, workflows, and distinct node types included. ComfyGPT provides the most comprehensive task coverage and dataset scope across both the training dataset and the benchmark.}
\label{tab:workflow_comparison}
\vspace{-4mm}
\end{table}

ComfyGPT comprises four specialized agents: ReformatAgent, FlowAgent, RefineAgent, and ExecuteAgent. 
ReformatAgent transforms JSON-formatted ComfyUI workflows into understandable workflow diagrams, emphasizing individual node connections during training. FlowAgent generates workflows based on task instructions using the GRPO algorithm, while also autonomously correcting errors during generation. RefineAgent integrates LLMs with retrieval-based knowledge, refining and validating workflows to ensure topological consistency. ExecuteAgent converts validated diagrams back into JSON formats compatible with the ComfyUI server for execution. As illustrated in Fig.~\ref{fig:intro}, ComfyGPT accurately and efficiently generates ComfyUI workflows aligned with user instructions across various tasks.

To train and evaluate ComfyGPT, we introduce \textbf{FlowDataset} and \textbf{FlowBench}. FlowDataset comprises $13,571$ examples of ComfyUI workflows paired with instruction prompts. These are divided into $6$ main categories with six additional subcategories, ensuring comprehensive coverage of image editing tasks. FlowBench, a standardized benchmark, includes $1,000$ entries for evaluation while the remaining 12,571 samples are reserved for training. FlowBench improves upon existing benchmarks with higher node complexity, greater diversity, and broader task coverage, providing a robust evaluation framework not only for ComfyGPT but also for other LLM-based workflow generation approaches. Additionally, we propose four effective evaluation metrics: Format Validation (FV), Pass Accuracy (PA), Pass Instruct Alignment (PIA), and Pass Node Diversity (PND). Experimental results demonstrate that ComfyGPT significantly outperforms prior LLM-based methods in workflow generation tasks.

In summary, our contributions can be summarized as follows:
\begin{itemize}
    \item We propose ComfyGPT, a self-optimizing multi-agent system designed to automatically and effectively generate ComfyUI workflows for a wide range of tasks based on user instructions. 
    \item ComfyGPT introduces a novel approach to workflow generation by focusing on individual node connections rather than generating entire node structures, which significantly enhances the ability of LLMs to capture and handle complex interdependencies within ComfyUI's topological workflows.
    \item We present FlowDataset, a large-scale, open-source training set, along with a comprehensive benchmark called FlowBench. FlowBench surpasses existing benchmarks in terms of workflow complexity, node diversity, and task coverage, offering a robust evaluation framework for both ComfyGPT and other LLM-based platforms.
\end{itemize}

\begin{figure*}[t]
  \centering
  \includegraphics[width=0.85\linewidth]{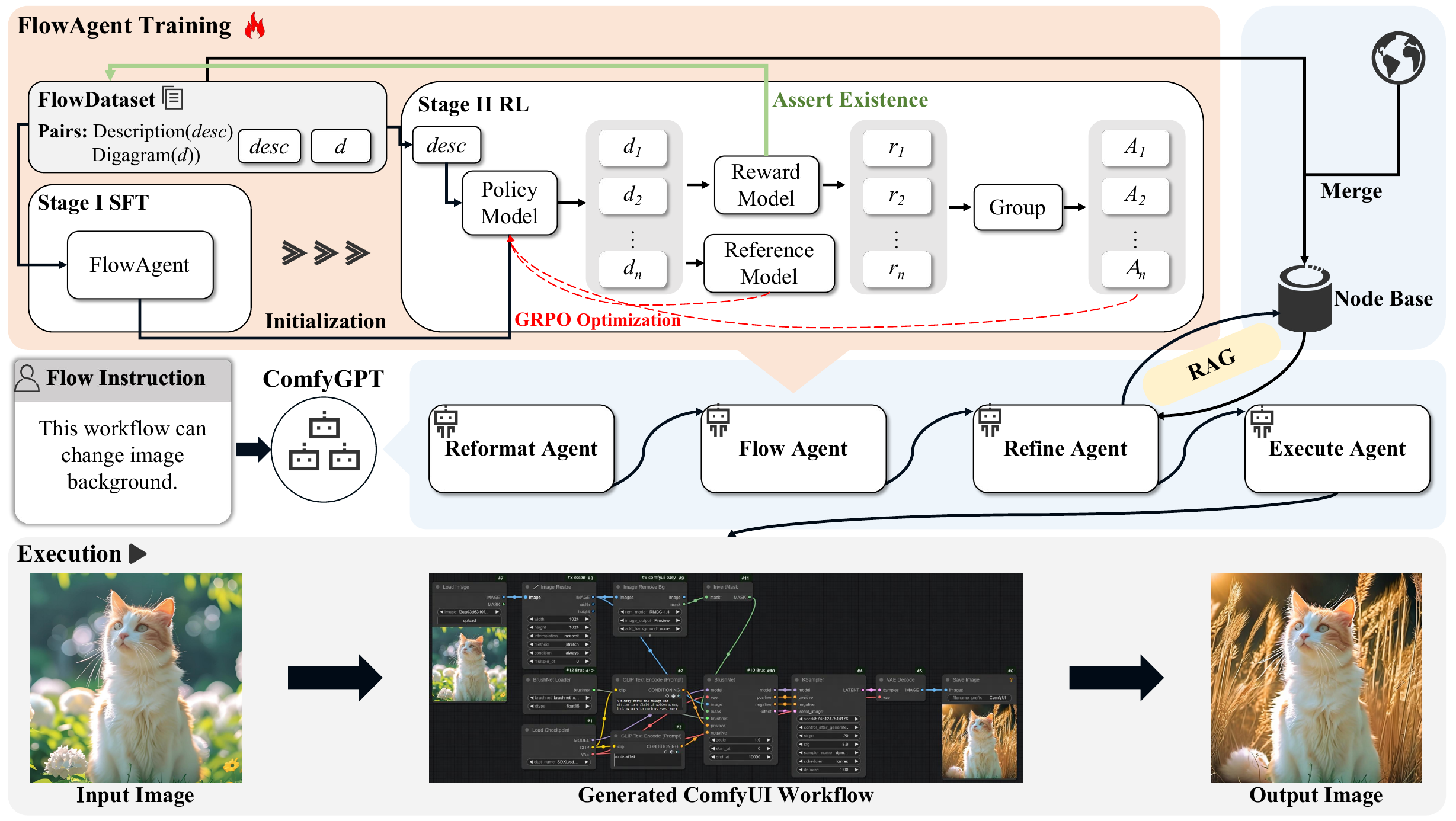}
  \caption{\textbf{Overview of the ComfyGPT pipeline for automated ComfyUI workflow generation.} Given a user instruction, ComfyGPT sequentially executes four specialized agents to generate workflows. ReformatAgent converts ComfyUI workflows into workflow diagrams during few-shot learning and FlowAgent training. FlowAgent generates workflow diagrams based on instructions, training with supervised fine-tuning (SFT) and optimized via GRPO. RefineAgent enhances the diagram quality by integrating LLMs with knowledge retrieval for validation and optimization. Finally, ExecuteAgent converts the optimized diagram into a ComfyUI-compatible JSON format and executes it within the ComfyUI environment.}
  \label{fig:pipe}
  \vspace{-4mm}
\end{figure*}

\section{Relate Work}
\label{sec:relate_work}
\subsection{Image Generation}

Image generation has achieved significant breakthroughs, demonstrating extensive application potential across various domains. Notably, text-to-image generation technology~\cite{DALL-E,DALL-E2,scaling,StableDiffusion,Imagen,gan-cls,df-gan,dm-gan,attngan,stackgan,stackgan++,glide,zhang2021cross,liu2024llm4genleveragingsemanticrepresentation}, such as Stable Diffusion~\cite{StableDiffusion}, allows users to generate rich and diverse images from simple textual descriptions. Building on this foundation, some methods~\cite{controlnet,gligen,t2i,densediff,layout-guidances,boxdiff,TraDiffusion,freecontrol,attention-refocusing,zero-spatial,loco}, like ControlNet~\cite{controlnet}, have introduced additional conditions to enhance the controllability and precision of image generation. More remarkable image generation tasks have become possible, such as style transfer~\cite{ip-adapter}, image restoration~\cite{zeng2021cr,xia2023diffir}, outpainting~\cite{sabini2018painting}, face swapping~\cite{ren2023pbidr,guo2021sample,gecer2021ostec,an_2021_pfc_iccvw,an_2022_pfc_cvpr,deng2018arcface,deng2018menpo,deng2020subcenter,Deng2020CVPR,guo2018stacked}, 3D image generation~\cite{boss2024sf3d,hunyuan3d22025tencent,mildenhall2021nerf} and video generation~\cite{cogvideo,animatediff}. 

\subsection{AI Agent with LLM}
LLM-based Agents are intelligent systems that leverage the powerful understanding and generation capabilities of large language models(LLM)~\cite{language,palm,training,opt,glm,llama,wei2022emergent,deepseekr1,genmac,mora,xu2025comfyui}, integrated with external tools to accomplish specific tasks. There are two main approaches to building LLM-based Agents. One involves full fine-tuning or parameter-efficient methods like LoRA, such as ComfyGen~\cite{comfygen}, a prompt-adaptive LLM-based agent that automatically generates ComfyUI workflow. However, ComfyGen is restricted to text-to-image tasks, failing to handle more complicated image generation tasks. The other leverages pre-trained LLM capabilities through few-shot learning~\cite{few-shot}, capitalizing on extensive corpus training while avoiding additional training overhead. For example, HuggingGPT~\cite{hugginggpt} leverages few-shot learning for intent interpretation and utilizes Huggingface API to perform diverse tasks, while ComfyBench~\cite{comfybench} employs a similar few-shot approach to create a multi-agent system that automatically generates ComfyUI workflows. However, these open-loop approaches often fail to fully utilize the potential of LLM as agents, being constrained by context limitations while introducing additional computational costs and latency during inference. Besides, they rely on multi-step few-shot learning of LLM, leading to error accumulation and suboptimal results.

\section{Method}
\subsection{ComfyUI Modeling}
ComfyUI is an open-source platform designed for creating and managing complex workflows, with particular strengths in image generation. It provides an intuitive interface for interacting with various deep learning models, enabling data processing and visualization. 

\begin{figure}[t]
  \centering
  \includegraphics[width=1\linewidth]{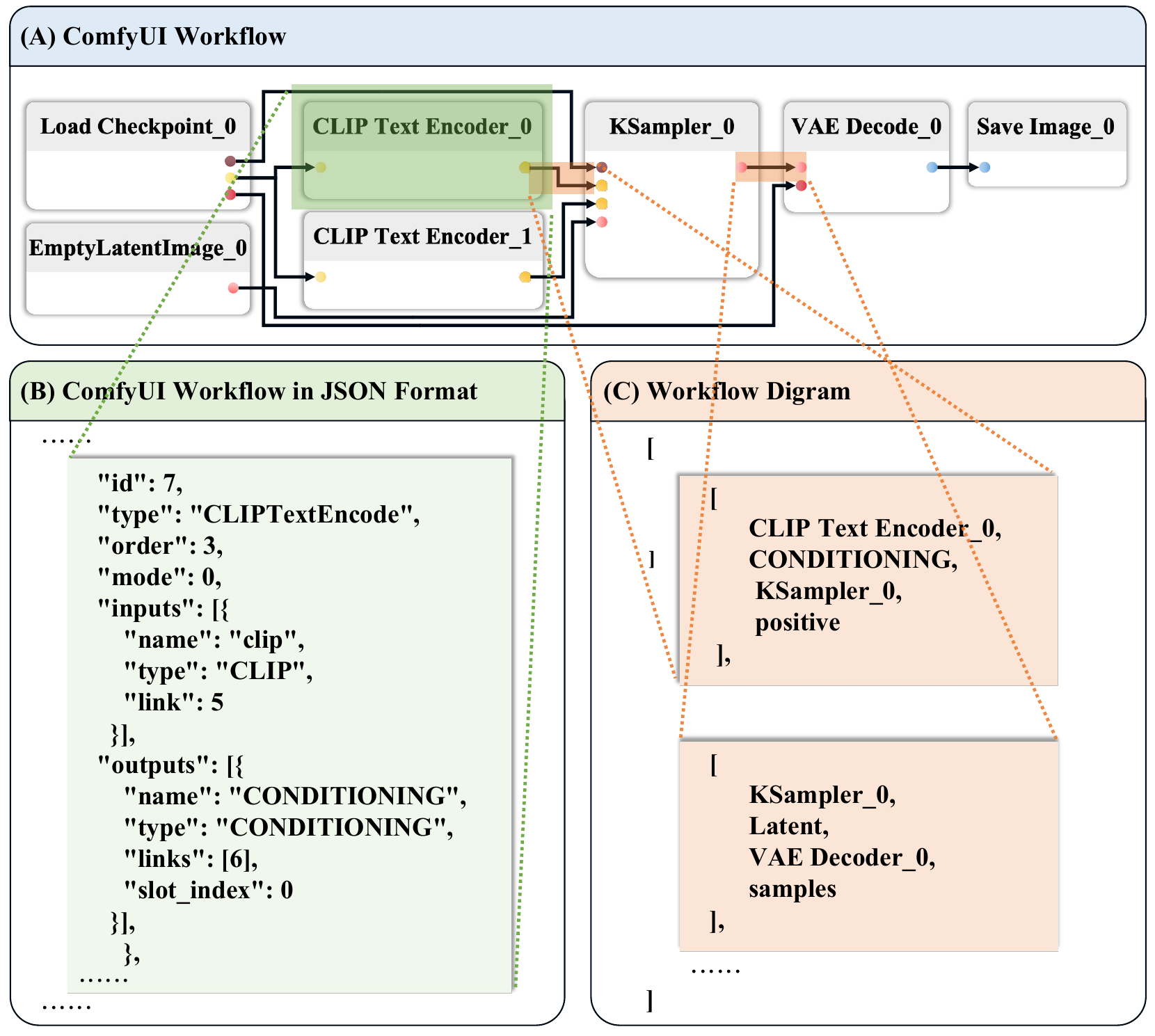}
  \caption{\textbf{Illustration of Different Representations of ComfyUI Workflows.} Instead of generating the entire JSON format ComfyUI workflows, we introduce a new workflow diagram to generate individual links between the processing nodes (C).}
  \label{fig:format}
  \vspace{-2mm}
\end{figure}

One of the key strengths of ComfyUI lies in its use of the JSON format to store node information, as shown in Fig.~\ref{fig:format}(B). This format allows the platform to easily interpret custom nodes and integrate them into the interface. Specifically, ComfyUI breaks the model inference process into modular components called nodes, represented as follows:
\begin{equation}
    \mathbb{N} = \{n_k\}, k = 1, 2, \dots, K,
\end{equation}
where $n_{k}$ represents an individual node, such as a model, tool, or specific function, as shown in Fig.~\ref{fig:format}(A). Each node $n_k$ has multiple logical inputs and outputs, defined as:
\begin{equation}
    \mathbb{I}^k = \{I^k_m\},
    \mathbb{O}^k =  \{O^k_m\},m=1,2,\dots,M,
\end{equation}
where $I^k_m$ denotes the  $m$-th input name of node $n_k$, $O^k_m$ represents the $m$-th output name of node $n_k$. These inputs and outputs are linked together to form topological workflow structures, enabling the execution of various tasks.

ComfyUI's modular and customizable design not only visualizes task workflows clearly but also provides users with extensive flexibility and control. However, this flexibility also brings challenges. The complex interconnections between nodes and the platform's diverse functionalities can make it difficult for beginners to get started.

\subsection{ComfyGPT}
We propose ComfyGPT, a self-optimizing and end-to-end multi-agent system designed to generate workflows for ComfyUI based on user-provided instructions.

ComfyGPT comprises four key agents: ReformatAgent, FlowAgent, RefineAgent, and ExecuteAgent. As shown in Fig.~\ref{fig:pipe}, the ReformatAgent extracts a workflow logic diagram from the ComfyUI JSON format during training or few-shot inference. Subsequently, the FlowAgent generates a logic diagram based on the processed flow description provided by the ReformatAgent. The RefineAgent refines the generated diagram using a node database to improve its accuracy. Finally, the ExecuteAgent traverses the refined diagram into a ComfyUI-compatible JSON workflow and executes it on the ComfyUI server.

\noindent\textbf{RefomatAgent.} As shown in Fig.~\ref{fig:format}(B), ComfyUI workflows are often lengthy and include redundant information, making them diffucult for LLMs to process due to the context length limitations. 
Furthermore, the complexity of node connections makes it challenging for LLM to effectively navigate and understand node relationships.

To address this, the ReformatAgent processes ComfyUI workflows in JSON format into simplified and more intuitive logic diagrams, represented as  $\mathbb{D}$ (illustrated in Fig.~\ref{fig:format} (C)) by specifically on the links between nodes. This transformation is defined as follows:
\begin{equation}
   \mathbb{D} = \{l_i\},i=1,2,...,I,
\end{equation}
where, each link $l_i$ is represented as:
\begin{equation}
   l_i = [ n_{out}, O^{out}_{j}, n_{in}, I^{in}_{k} ], j\in M,k\in M,
   \label{eq:link}
\end{equation}
where $n_{out}$ and $n_{in}$ represent the output and input nodes, while $O^{out}_{j}$ and $I^{in}_{k}$ denote the specific outputs and inputs being connected.
For example, the first box shown in Fig.~\ref{fig:format}(C) illustrates a link $l_i$, where \textit{CLIP Text Encoder\_0}, \textit{CONDITIONING} correspond to  $n_{out}$ and $O^{out}_{j}$, respectively, and \textit{KSampler\_0}, \textit{positive} correspond to $n_{in}$ and $I^{in}_{k}$. To ensure clarity and sequence, nodes with identical names are numerically differentiated, such as \textit{CLIP Text Encoder\_0} and \textit{CLIP Text Encoder\_1}, as shown in Fig.~\ref{fig:format}(A). This numbering ensures accurate representation of links and their order in the workflow.

\noindent\textbf{FlowAgent.} 
While large language models (LLMs) demonstrate strong few-shot capabilities, generating ComfyUI workflows remains a significant challenge due to their abstract structure. Open-loop methods often fail because LLMs lack exposure to similar training data. A common solution is using multi-agent systems to break down complex workflows, but this exacerbates errors due to LLMs' context length limitations, often resulting in deviations in the final output.

To address these issues, we propose the FlowAgent, a self-optimizing workflow generator powered by LLMs. The FlowAgent is trained using supervised fine-tuning (SFT) and reinforcement learning (RL) to improve workflow generation accuracy while enabling self-correction and iterative optimization. As shown in Fig.~\ref{fig:pipe}, in the SFT stage, each training instance consists of a workflow description $desc$ and workflow diagram $d$. The objective function for training is:

\begin{equation}
\resizebox{0.9\hsize}{!}{$
\mathcal{J}_{SFT}(\theta) = \mathbb{E}_{\text{desc}, d \sim P(FD)}
\left[ \sum_{t=1}^{T}  \log \pi_\theta(d_{,t} = i \mid \text{desc}, d_{,<t}) \right],
$}
\end{equation}
where $\pi_\theta$ is the trainable backbone of FlowAgent, $d$ represents the generated workflow diagram. $d_{,<t}$ refers to the sequence of tokens generated before time step $t$, $\pi_\theta(d_{,t} = i \mid \text{desc}, d_{,<t})$ is the probability that the LLM predicts token $d_{,t}$ in step $t$.

Training with workflow diagrams enables FlowAgent to effectively generate accurate diagrams from natural language descriptions. However, hallucination errors—where nodes not present in the workflow are fabricated—remain a challenge at this stage. To mitigate hallucination issues and further refine its outputs, FlowAgent undergoes reinforcement learning using the GRPO algorithm~\cite{deepseekmath}. The RL objective function is defined as:
\begin{equation}
\begin{split}
\mathcal{J}_{GRPO}(\theta) = \mathbb{E} \big[ desc \sim P(FD), \{d_i\}_{i=1}^{G} \sim \pi_{\theta_{\text{old}}} (D|desc) \big] \\
\frac{1}{G} \sum_{i=1}^{G} \frac{1}{|d_i|} \sum_{t=1}^{|d_i|} \Bigg\{ \min \bigg[ \frac{\pi_{\theta}(d_{i,t} | desc, d_{i,<t})}{\pi_{\theta_{\text{old}}}(_{i,t} | desc, d_{i,<t})} \hat{A}_{i,t},\\ 
\operatorname{clip} \left( \frac{\pi_{\theta}(d_{i,t} | desc, d_{i,<t})}{\pi_{\theta_{\text{old}}}(d_{i,t} | desc, d_{i,<t})}, 1-\varepsilon, 1+\varepsilon \right) \hat{A}_{i,t} \bigg] \\
- \beta \mathbb{D}_{KL} \big[ \pi_{\theta} \| \pi_{\text{ref}} \big] \Bigg\},
\end{split}
\label{eq:loss}
\end{equation}
where $\pi_{\theta}$ and $\pi_{\theta_{\text{old}}}$ denote the current and previous policy models initialized after the SFT stage, while $\pi_{\text{ref}}$serves as a reference model for KL penalty. $G$, $\beta$, and $\varepsilon$ are hyper-parameters used for group computation, clipping, and the KL penalty, respectively. $d_i$ represents $i$-th generated workflow diagram in the group $G$.The advantage $\hat{A}_{i,t}$ is calculated based on the relative rewards, defined as:
\begin{equation}
    \hat{A}_{i,t} = \frac{r_i-mean(r)}{std(r)}
\end{equation}
where $r = \{r_1, r_2, ... r_G\}$ are reward values produced by reward model for $d = \{d_1, d_2, ... d_G\}$ in one group computation. In our task, we introduce a new reward model to support GRPO training, which is defined as:
\begin{equation}
r_i =
\begin{cases}
0 & \text{if } \exists n_j \in d_i \text{ such that } n_j \notin N^T \\
1 & \text{if } \forall n_j \in d_i, n_j \in N^T
\end{cases}
\end{equation}
where $n_j$ is the node name from $d_i$, $N^T$ is the collection of node names. This guarantees the generation of workflow diagrams free of fictitious nodes while enhancing compliance with valid node structures.

By integrating SFT and RL, FlowAgent achieves exceptional accuracy in generating workflows across a wide variety of tasks, overcoming challenges such as hallucinations and context limitations. This optimization process ensures robust and reliable performance in modular workflow generation.

\noindent\textbf{RefineAgent.} 
Despite improvements from SFT and GRPO training, there may still be cases of outdated information on nodes. For instance, when the name of a ComfyUI node is changed in an updated GitHub repository (e.g., 'Text Box' becoming 'LayerUtility: TextBox'), retraining the model with newly collected data would be a significant challenge.

Therefore, we develop the RefineAgent by integrating a large language model (LLM) with knowledge retrieval capabilities. The RefineAgent functions as a secondary inspection and correction mechanism following the self-correction of the FlowAgent, serving as a protective barrier within the entire system.

To ensure the system remains up-to-date, we maintain a continuously updated node database $\mathcal{K}$ from the Internet, containing $6,362$ unique nodes. Each node's collection includes the node name along with input and output specifications, detailing their specific types and names. To retrieve nodes from $\mathcal{K}$, we use a language model to encode the node names into semantic embedding vectors, which can be defined as:
\begin{equation}
   \vec{e_i} =E_{\theta}(n_i),
\end{equation}
where $e_i$ represents the embedding vector of the node's name $n_i$. 
We measure the similarity between the two nodes \( n_i \) and \( n_j \) by calculating the cosine similarity between their embedding vectors, which can be formulated as follows:
\begin{equation}
\text{s}(n_i, n_j) = \frac{\vec{e_i} \cdot \vec{e_j}}{\|\vec{e_i}\| \|\vec{e_j}\|},
\end{equation}
where $s(n_i, n_j)$ represents the similarity score between nodes $n_i$ and $n_j$. For an incorrect node $n_{ic}$—a node that fails to execute in the current ComfyUI environment—we calculate a similarity score list $\mathbf{S}_{ic}$ comparing $n_{ic}$ to other nodes in the database $\mathcal{K}$.

Subsequently, we select the top $k$ most similar nodes and retrieve their information from the node database, which can be defined as follows: 
\begin{equation}
\operatorname{TopK}(n_{ic}, k)=\{ K(v_j) \mid v_j \in \operatorname{argsort}(\mathbf{S}_{ic})[-k:] \},
\end{equation}
where \( \operatorname{argsort}(\mathbf{S}_{ic})[-k:] \) ranks the top $k$ most similar nodes and \( K(v_j) \) retrieves detailed information about the corresponding node indexed by \( v_j \). Using the workflow diagram $d$ generated by FlowAgent, the user's input description $desc$, the incorrect node $n_{ic}$, and the top $k$ candidate nodes $\operatorname{TopK}(n_{ic}, k)$, LLM selects the most suitable candidate node $n_c$ from $\mathcal{K}$,
\begin{equation}
    n_c =LLM(d, desc, n_i, \operatorname{TopK}(n_{ic}, k)).
\end{equation}
Finally, the incorrect node $n_{ic}$ is replaced with the corrected node $n_c$. More implementation details can be found in the Appendix.

\noindent\textbf{ExecuteAgent.}
The ExecuteAgent completes the process by converting the refined workflow diagram (Fig.~\ref{fig:format}(C)) back into ComfyUI-compatible JSON format (Fig.~\ref{fig:format}(B)). This conversion reverses the transformation performed by the ReformatAgent. Finally, the ExecuteAgent uploads the JSON-formatted workflow to the ComfyUI server, enabling the system to generate the desired output in response to the user's instructions.

\section{Data Construction}

\begin{figure}[t]
  \centering
  \includegraphics[width=1\linewidth]{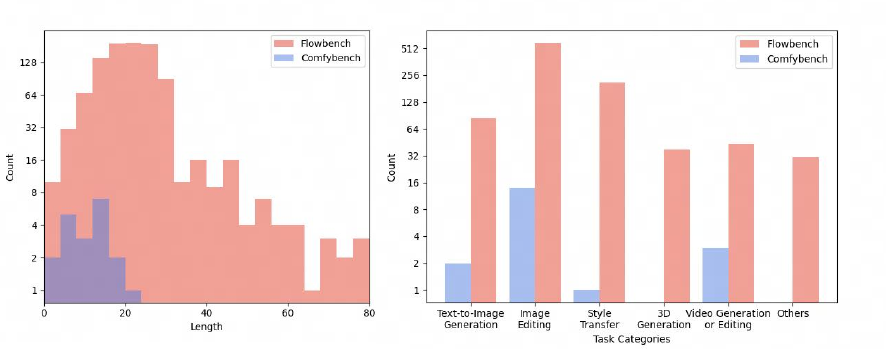}
  \caption{\textbf{Illustration of length (left) and task categories (right) distribution in FlowBench.} The length is calculated by the number of nodes contained in each workflow.}
  \label{fig:bench_2}
  \vspace{-2mm}
\end{figure}

\begin{figure*}[t]
  \centering
  \includegraphics[width=0.85\linewidth]{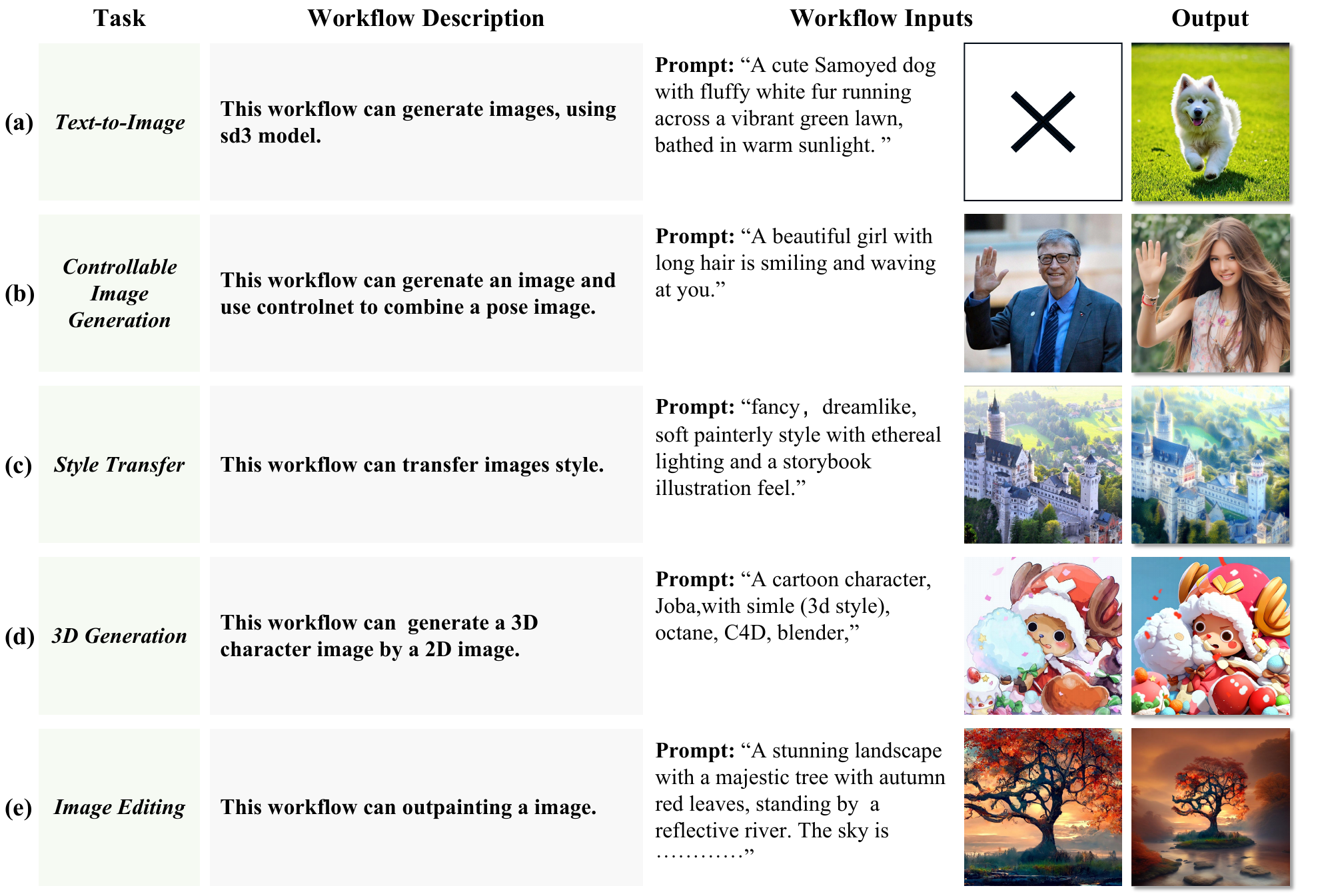}
  \caption{\textbf{Qualitative results of ComfyGPT across different tasks.} We place the specific workflows for the examples from (a) to (e) in the Appendix.}
  \label{fig:ability}
  \vspace{-2mm}
\end{figure*}

\subsection{FlowDataset} 
To train the FlowAgent, we create a large dataset called \textbf{FlowDataset}, containing $13,571$ workflows paired with corresponding instructions. These are organized into $6$ core categories and $6$ subcategories, making FlowDataset the most comprehensive workflow dataset available.

\noindent\textbf{Collection and revise.} FlowDataset is curated from a large volume of workflow-description pairs collected from ComfyUI community. However, the raw data includes significant noise. We first filter out the workflows that fail to meet ComfyUI's execution standards and remove irrelevant information—such as "Reroute" nodes—to simplify and optimize the training data for LLMs. Additionally, we refine and clarify chaotic user-written descriptions using ChatGPT-4.0 Mini and further eliminate noisy data for improved consistency. Details of the data refinement process can be found in the Appendix.

\noindent\textbf{Categorization.} 
We classify the dataset into $6$ broad categories using ChatGPT-4.0 Mini for semantic analysis and summarization of workflow descriptions. The core categories include: Text-to-Image Generation, Image Editing, Style Transfer, 3D Generation, Video Editing or Generation, and Others. Given ComfyUI's focus on image processing, we further subdivide \textbf{Image Editing} into $6$ subcategories: HD Upscaling/Image Restoration, Redrawing, Outpainting, Character-Based Guidance, Face Swap, and Background Change/Remove.

\noindent\textbf{Classification.} Using the robust capabilities of LLMs, we classify and organize entries into these categories. Each data sample includes a natural language description, its corresponding JSON-formatted workflow, and the associated category labels, resulting in $13,571$ high-quality data entries. Additional details on the construction and structure of FlowDataset are provided in the Appendix.

\subsection{FlowBench}
\noindent\textbf{Benchmark Dataset.} To evaluate the performance of ComfyGPT, we partition $1,000$ samples from FlowDataset to create a specialized test set called \textbf{FlowBench}. As shown in Tab.~\ref{tab:workflow_comparison}, FlowBench is more comprehensive than existing benchmarks, covering a broader range of categories and including data types for image, video, and 3D generation. It surpasses its predecessors in diversity and scale, none of which include this full range of functionality. Fig.~\ref{fig:bench_2} shows the distribution of workflow lengths and categories, demonstrating that FlowBench outperforms ComfyBench in terms of both the average length of each workflow and the number of workflows across different categories. 
FlowBench is not only useful for evaluating ComfyGPT but can also serve as a general benchmark for assessing the task-solving abilities of LLMs. Detailed information about FlowBench is provided in the Appendix.

\noindent\textbf{Evaluation Metrics.} 
To comprehensively evaluate ComfyGPT, we propose four key metrics: Format Validation (FV), Pass Accuracy (PA), Pass Instruct Alignment (PIA), and Pass Node Diversity (PND). \textbf{FV} measures whether the generated workflows are in the correct format.
\textbf{PA} calculates the percentage of generated workflows that can be successfully executed on the ComfyUI server.
\textbf{PIA} evaluates how well the executed workflows align with the given user instructions.
\textbf{PND} counts the number of unique node types present in the successfully executed workflows, reflecting the model's ability to generate diverse and varied workflows.


\section{Experiment}
\subsection{Experiment Settings}
\label{sec:experiment}

\noindent{\textbf{Baseline.}} We evaluate our approach against two multi-agent methods: ComfyBench~\cite{comfybench}, a specialized workflow-generation method for ComfyUI, and Multi-Agent (Debate)~\cite{muti-agent-debate}, a general debate-based multi-agent framework. Additionally, we compare with four state-of-the-art (SOTA) closed-source models for few-shot evaluation: ChatGPT-4-32~\cite{gpt-4}, ChatGPT-4o~\cite{gpt-4}, Claude-3-5-sonnet~\cite{Claude}, and Claude-3-7-sonnet~\cite{Claude3-7}.
We also evaluate three different open-source models: Llama-13B~\cite{llama}, Baichuan2-13B-Base~\cite{baichuan2}, and Chatglm3-6B~\cite{glm2024chatglm}, used as foundational models to assess the effectiveness of our approach.

\noindent{\textbf{Implementation Details.}} We use Qwen2.5-14B-Base~\cite{qwen2.5} as the backbone for FlowAgent in ComfyGPT.
The SFT training is performed on $4$ NVIDIA A100 GPUs (80GB) over $3$ epochs. For the RL stage, $8$ NVIDIA A100 GPUs (80GB) are used to train the model over a total of $300$ steps. The RefineAgent utilizes the Qwen2.5-14B-Instruct model as its base for refining workflow diagrams. It also employs the Multilingual-MiniLM-L12-v2 model to compute semantic embeddings, with the top $k=5$ most similar nodes being retrieved. Additional implementation details can be found in the Appendix.

\begin{table}[t]  
    
    \centering

    \resizebox{\columnwidth}{!}{ \begin{tabular}{lccccc}
    \toprule
    \textbf{Method} &\textbf{Venues}&  \textbf{FV(\%)$\uparrow$}& \textbf{PA(\%)$\uparrow$}& \textbf{PIA(\%)$\uparrow$}& \textbf{PND$\uparrow$} \\

    \midrule
    ChatGPT-4-32&preprint 2024&  11.2& 11.0&11.0& 67\\
    ChatGPT-4o&preprint 2024 & 12.6& 12.4&12.4& 59\\
    Claude-3-5-sonnet&preprint 2024 & 12.6& 12.5&12.5& 66\\
    Claude-3-7-sonnet&preprint 2024 & 16.8& 15.8&15.8& 72\\

    \midrule
    Muti-Agent(debate)&ICML 2024&20.2&17.3&15.9& 88\\
    ComfyAgent&CVPR 2025&15.2& 14.5&14.2&50\\
    
    \midrule
    
    Llama-13B&preprint 2023 & 87.0& 85.2&84.2& 251\\
    Baichuan2-13B-Base&preprint 2023 & 87.2& 84.4&82.9& 294\\
    Chatglm3-6B&preprint 2024 & 87.2& 84.3& 82.9& 227\\
   \midrule
    ComfyGPT&-&\textbf{90.0}&\textbf{86.0}&\textbf{84.6}&\textbf{320}\\
    
    \bottomrule
    \end{tabular}
    }
   \caption{\textbf{Quantitative comparison of model performance on the proposed FlowBench.} ComfyGPT demonstrates superior performance across all metrics, establishing its effectiveness compared to baseline and state-of-the-art models. The model with
the best performance is in bold.}

    \label{table:baseline}
    \vspace{-2.5mm}
\end{table}
    
      

    
    

\begin{table}[t]  
    
    \centering
    \scalebox{0.8}{ \begin{tabular}{c|ccc}
    \toprule
    \textbf{Method $\xrightarrow[]{}$}&Muti-Agent(debate)&ComfyAgent&ComfyGPT\\
      
    \midrule
    \textbf{Pass(\%)$\uparrow$}&52.0&56.0& \textbf{81.0}\\

    \bottomrule
    \end{tabular}
    }
    \caption{\textbf{Quantitative comparison result on ComfyBench.} ComfyGPT demonstrates the superior performance of ComfyGPT with a significant improvement in Pass Rate compared with other methods.}
    \label{table:cmp}
    \vspace{-2.5mm}
\end{table}
\begin{table}[t]  
    \centering
    \resizebox{\columnwidth}{!}{\begin{tabular}{ccccc|cccc}
    \toprule
    \multicolumn{5}{c|}{\textbf{ComfyGPT}} & \multicolumn{4}{c}{\textbf{FlowBench}}  \\ \cmidrule(lr){1-5} \cmidrule(lr){6-9} 
     \textbf{FlowAgent}&\textbf{ReformatAgent}&\textbf{ExecuteAgent}& \textbf{w GRPO}& \textbf{RefineAgent}& \textbf{FV(\%)$\uparrow$}& \textbf{PA(\%)$\uparrow$}& \textbf{PIA(\%)$\uparrow$}& \textbf{PND$\uparrow$}\\
     
    \midrule
   \checkmark & $\times$& $\times$& $\times$& $\times$&74.8&66.0&64.2&183\\
   \checkmark & \checkmark& \checkmark& $\times$& $\times$& 85.9& 83.5& 82.4&283\\
   \checkmark & \checkmark& \checkmark& \checkmark& $\times$& 87.0& 84.7& 83.5& 275\\

   \checkmark & \checkmark& \checkmark& \checkmark& \checkmark & \textbf{90.0}& \textbf{86.0}& \textbf{84.6}& \textbf{320}\\
    \bottomrule
    
    \end{tabular}}
     \caption{\textbf{Quantitative ablation result of different agents in ComfyGPT.} The use of GRPO improves FV, PA, and PIA metrics. Each component of ComfyGPT plays a significant role in enhancing the overall performance.}
    \label{table:abl}
    \vspace{-2mm}
\end{table}
\noindent{\textbf{Evaluation Benchmark and Metrics}} Our evaluation is conducted both on FlowBench and ComfyBench~\cite{comfybench}. In FlowBench, we assess four metrics: Format Validation (FV), Pass Accuracy (PA), Pass Instruct Alignment (PIA), and Pass Node Diversity (PND). In ComfyBench, we evaluate the Pass Rate metric, which is similar to the Pass Accuracy metric in FlowBench.

\subsection{Quantitative Evaluation}
\noindent{\textbf{On benchmark FlowBench.}} As shown in Tab.~\ref{table:baseline}, ComfyGPT achieves substantial improvements over all baselines across all metrics: FA ($90.0\%$), PA ($86.0\%$), PIA (84.6$\%$), and PND ($320$). The closed-source methods rely on constrained information and pre-trained capabilities, which hampers their ability to effectively learn node information and construct workflow diagrams based on feedback, leading to lower performance. For example, their best results are limited to an FV of $16.8\%$ and a PA of $15.8\%$. ComfyAgent and Multi-agent(Debate) generate ComfyUI workflow by constructing multi-agent systems. Similarly, they struggle with task understanding and output generation, as LLMs haven't encountered similar training data. Additionally, during the multi-agent inference process, it becomes susceptible to error accumulation, leading to biased outcomes. Thus, their best performance is FV $20.2\%$ and PA $17.3\%$.
The marked increase in accuracy further suggests that ComfyGPT's self-optimizing multi-agent system is the key factor in its superior performance in understanding and responding to user instructions.

\noindent{\textbf{On benchmark ComfyBench.}}
As shown in Tab.~\ref{table:cmp},  our proposed ComfyGPT achieves state-of-the-art performance on cross-benchmark evaluation, where it leads by $25\%$.

\subsection{Qualitative Evaluation}
As illustrated in Fig.~\ref{fig:ability}, ComfyGPT showcases its strong capability to generate diverse workflows from simple instructions. Its flexibility extends beyond common tasks like text-to-image generation, controllable image creation, and style transfer. Furthermore, ComfyGPT demonstrates an understanding of specific model usage. For example, in Fig.~\ref{fig:ability}(a), it successfully generates a text-to-image (T2I) workflow incorporating the SD3 model when given the instruction "using sd3 model." This highlights its ability to adapt workflows to align with user-specified requirements.
\subsection{Ablation Study} 
As shown in Tab.~\ref{table:abl}, each agent contributes to notable improvements across all metrics. Specifically, the ReformatAgent and ExecuteAgent yield significant enhancements of $11.1\%$ in Format Validation(FV) and $17.5\%$ in Pass Accuracy(PA), respectively. By structuring workflows into diagrams, the ReformatAgent enables the LLM to better understand node connections, improving its workflow comprehension. In addition, using the GRPO algorithm leads to incremental gains of $1.1\%$ in FV and $1.2\%$ in PA, demonstrating its effectiveness in refining workflows through reinforcement learning (RL). We also observe a decrease in PND, which is a normal occurrence as the model gradually converges during the RL training process, leading to a reduction in diversity. This is a positive outcome that indicates improved stability in the model's performance. 

\section{Conclusion}
In summary, this paper introduces ComfyGPT, a self-optimizing multi-agent system designed to automatically generate ComfyUI workflows from task descriptions with higher accuracy and flexibility. By focusing on creating individual node links rather than entire workflows at once, ComfyGPT achieves more precise and adaptable workflow generation. With the help of reinforcement learning (RL), it continuously improves its performance over iterations. Additionally, we provide FlowDataset, a comprehensive collection of workflows, and FlowBench, a benchmark for evaluating workflow generation methods. Experimental results demonstrate that ComfyGPT surpasses existing LLM-based approaches in all key metrics, highlighting its effectiveness and adaptability.

\clearpage
\bibliography{aaai2026}

\begin{thebibliography}{73}
\providecommand{\natexlab}[1]{#1}

\bibitem[{An et~al.(2022)An, Deng, Guo, Feng, Zhu, Jing, and Tongliang}]{an_2022_pfc_cvpr}
An, X.; Deng, J.; Guo, J.; Feng, Z.; Zhu, X.; Jing, Y.; and Tongliang, L. 2022.
\newblock Killing Two Birds with One Stone: Efficient and Robust Training of Face Recognition CNNs by Partial FC.
\newblock In \emph{CVPR}.

\bibitem[{An et~al.(2021)An, Zhu, Gao, Xiao, Zhao, Feng, Wu, Qin, Zhang, Zhang, and Fu}]{an_2021_pfc_iccvw}
An, X.; Zhu, X.; Gao, Y.; Xiao, Y.; Zhao, Y.; Feng, Z.; Wu, L.; Qin, B.; Zhang, M.; Zhang, D.; and Fu, Y. 2021.
\newblock Partial FC: Training 10 Million Identities on a Single Machine.
\newblock In \emph{ICCVW}.

\bibitem[{Anthropic(2024{\natexlab{a}})}]{Claude}
Anthropic. 2024{\natexlab{a}}.
\newblock Claude 3 Haiku: Our Fastest Model Yet.
\newblock Accessed: 2024-10-15.

\bibitem[{Anthropic(2024{\natexlab{b}})}]{Claude3-7}
Anthropic. 2024{\natexlab{b}}.
\newblock Claude 3.7 Sonnet and Claude Code.
\newblock Accessed: 2024-10-15.

\bibitem[{Baichuan(2023)}]{baichuan2}
Baichuan. 2023.
\newblock Baichuan 2: Open Large-scale Language Models.
\newblock \emph{arXiv preprint arXiv:2309.10305}.

\bibitem[{Boss et~al.(2024)Boss, Huang, Vasishta, and Jampani}]{boss2024sf3d}
Boss, M.; Huang, Z.; Vasishta, A.; and Jampani, V. 2024.
\newblock Sf3d: Stable fast 3d mesh reconstruction with uv-unwrapping and illumination disentanglement.
\newblock \emph{arXiv preprint arXiv:2408.00653}.

\bibitem[{Brown et~al.(2020{\natexlab{a}})Brown, Mann, Ryder, Subbiah, Kaplan, Dhariwal, Neelakantan, Shyam, Sastry, Askell et~al.}]{language}
Brown, T.; Mann, B.; Ryder, N.; Subbiah, M.; Kaplan, J.~D.; Dhariwal, P.; Neelakantan, A.; Shyam, P.; Sastry, G.; Askell, A.; et~al. 2020{\natexlab{a}}.
\newblock Language models are few-shot learners.
\newblock \emph{Advances in neural information processing systems}, 33: 1877--1901.

\bibitem[{Brown et~al.(2020{\natexlab{b}})Brown, Mann, Ryder, Subbiah, Kaplan, Dhariwal, Neelakantan, Shyam, Sastry, Askell et~al.}]{few-shot}
Brown, T.; Mann, B.; Ryder, N.; Subbiah, M.; Kaplan, J.~D.; Dhariwal, P.; Neelakantan, A.; Shyam, P.; Sastry, G.; Askell, A.; et~al. 2020{\natexlab{b}}.
\newblock Language models are few-shot learners.
\newblock \emph{Advances in neural information processing systems}, 33: 1877--1901.

\bibitem[{Chen, Laina, and Vedaldi(2024)}]{layout-guidances}
Chen, M.; Laina, I.; and Vedaldi, A. 2024.
\newblock Training-free layout control with cross-attention guidance.
\newblock In \emph{Proceedings of the IEEE/CVF Winter Conference on Applications of Computer Vision}, 5343--5353.

\bibitem[{Chowdhery et~al.(2023)Chowdhery, Narang, Devlin, Bosma, Mishra, Roberts, Barham, Chung, Sutton, Gehrmann et~al.}]{palm}
Chowdhery, A.; Narang, S.; Devlin, J.; Bosma, M.; Mishra, G.; Roberts, A.; Barham, P.; Chung, H.~W.; Sutton, C.; Gehrmann, S.; et~al. 2023.
\newblock Palm: Scaling language modeling with pathways.
\newblock \emph{Journal of Machine Learning Research}, 24(240): 1--113.

\bibitem[{Couairon et~al.(2023)Couairon, Careil, Cord, Lathuiliere, and Verbeek}]{zero-spatial}
Couairon, G.; Careil, M.; Cord, M.; Lathuiliere, S.; and Verbeek, J. 2023.
\newblock Zero-shot spatial layout conditioning for text-to-image diffusion models.
\newblock In \emph{Proceedings of the IEEE/CVF International Conference on Computer Vision}, 2174--2183.

\bibitem[{DeepSeek-AI(2025)}]{deepseekr1}
DeepSeek-AI. 2025.
\newblock DeepSeek-R1: Incentivizing Reasoning Capability in LLMs via Reinforcement Learning.
\newblock arXiv:2501.12948.

\bibitem[{Deng et~al.(2020{\natexlab{a}})Deng, Guo, Liu, Gong, and Zafeiriou}]{deng2020subcenter}
Deng, J.; Guo, J.; Liu, T.; Gong, M.; and Zafeiriou, S. 2020{\natexlab{a}}.
\newblock Sub-center ArcFace: Boosting Face Recognition by Large-scale Noisy Web Faces.
\newblock In \emph{Proceedings of the IEEE Conference on European Conference on Computer Vision}.

\bibitem[{Deng et~al.(2019)Deng, Guo, Niannan, and Zafeiriou}]{deng2018arcface}
Deng, J.; Guo, J.; Niannan, X.; and Zafeiriou, S. 2019.
\newblock ArcFace: Additive Angular Margin Loss for Deep Face Recognition.
\newblock In \emph{CVPR}.

\bibitem[{Deng et~al.(2020{\natexlab{b}})Deng, Guo, Ververas, Kotsia, and Zafeiriou}]{Deng2020CVPR}
Deng, J.; Guo, J.; Ververas, E.; Kotsia, I.; and Zafeiriou, S. 2020{\natexlab{b}}.
\newblock RetinaFace: Single-Shot Multi-Level Face Localisation in the Wild.
\newblock In \emph{CVPR}.

\bibitem[{Deng et~al.(2018)Deng, Roussos, Chrysos, Ververas, Kotsia, Shen, and Zafeiriou}]{deng2018menpo}
Deng, J.; Roussos, A.; Chrysos, G.; Ververas, E.; Kotsia, I.; Shen, J.; and Zafeiriou, S. 2018.
\newblock The Menpo benchmark for multi-pose 2D and 3D facial landmark localisation and tracking.
\newblock \emph{IJCV}.

\bibitem[{Du et~al.(2023)Du, Li, Torralba, Tenenbaum, and Mordatch}]{muti-agent-debate}
Du, Y.; Li, S.; Torralba, A.; Tenenbaum, J.~B.; and Mordatch, I. 2023.
\newblock Improving factuality and reasoning in language models through multiagent debate.
\newblock In \emph{Forty-first International Conference on Machine Learning}.

\bibitem[{Gal et~al.(2024)Gal, Haviv, Alaluf, Bermano, Cohen-Or, and Chechik}]{comfygen}
Gal, R.; Haviv, A.; Alaluf, Y.; Bermano, A.~H.; Cohen-Or, D.; and Chechik, G. 2024.
\newblock Comfygen: Prompt-adaptive workflows for text-to-image generation.
\newblock \emph{arXiv preprint arXiv:2410.01731}.

\bibitem[{Gecer, Deng, and Zafeiriou(2021)}]{gecer2021ostec}
Gecer, B.; Deng, J.; and Zafeiriou, S. 2021.
\newblock OSTeC: One-Shot Texture Completion.
\newblock In \emph{Proceedings of the IEEE/CVF Conference on Computer Vision and Pattern Recognition (CVPR)}.

\bibitem[{GLM(2024)}]{glm2024chatglm}
GLM, T. 2024.
\newblock ChatGLM: A Family of Large Language Models from GLM-130B to GLM-4 All Tools.
\newblock arXiv:2406.12793.

\bibitem[{Guo et~al.(2021)Guo, Deng, Lattas, and Zafeiriou}]{guo2021sample}
Guo, J.; Deng, J.; Lattas, A.; and Zafeiriou, S. 2021.
\newblock Sample and Computation Redistribution for Efficient Face Detection.
\newblock \emph{arXiv preprint arXiv:2105.04714}.

\bibitem[{Guo et~al.(2018)Guo, Deng, Xue, and Zafeiriou}]{guo2018stacked}
Guo, J.; Deng, J.; Xue, N.; and Zafeiriou, S. 2018.
\newblock Stacked Dense U-Nets with Dual Transformers for Robust Face Alignment.
\newblock In \emph{BMVC}.

\bibitem[{Guo et~al.(2023)Guo, Yang, Rao, Liang, Wang, Qiao, Agrawala, Lin, and Dai}]{animatediff}
Guo, Y.; Yang, C.; Rao, A.; Liang, Z.; Wang, Y.; Qiao, Y.; Agrawala, M.; Lin, D.; and Dai, B. 2023.
\newblock Animatediff: Animate your personalized text-to-image diffusion models without specific tuning.
\newblock \emph{arXiv preprint arXiv:2307.04725}.

\bibitem[{Han et~al.(2024)Han, Liu, Jiang, Yan, Zhang, Yuan, Peng, and Liu}]{Infinity}
Han, J.; Liu, J.; Jiang, Y.; Yan, B.; Zhang, Y.; Yuan, Z.; Peng, B.; and Liu, X. 2024.
\newblock Infinity: Scaling Bitwise AutoRegressive Modeling for High-Resolution Image Synthesis.
\newblock arXiv:2412.04431.

\bibitem[{Hong et~al.(2022)Hong, Ding, Zheng, Liu, and Tang}]{cogvideo}
Hong, W.; Ding, M.; Zheng, W.; Liu, X.; and Tang, J. 2022.
\newblock Cogvideo: Large-scale pretraining for text-to-video generation via transformers.
\newblock \emph{arXiv preprint arXiv:2205.15868}.

\bibitem[{Huang et~al.(2024)Huang, Huang, Ning, Lin, Wang, and Liu}]{genmac}
Huang, K.; Huang, Y.; Ning, X.; Lin, Z.; Wang, Y.; and Liu, X. 2024.
\newblock GenMAC: Compositional Text-to-Video Generation with Multi-Agent Collaboration.
\newblock \emph{arXiv preprint arXiv:2412.04440}.

\bibitem[{Kim et~al.(2023)Kim, Lee, Kim, Ha, and Zhu}]{densediff}
Kim, Y.; Lee, J.; Kim, J.-H.; Ha, J.-W.; and Zhu, J.-Y. 2023.
\newblock Dense text-to-image generation with attention modulation.
\newblock In \emph{Proceedings of the IEEE/CVF International Conference on Computer Vision}, 7701--7711.

\bibitem[{Kirillov et~al.(2023)Kirillov, Mintun, Ravi, Mao, Rolland, Gustafson, Xiao, Whitehead, Berg, Lo et~al.}]{kirillov2023segment}
Kirillov, A.; Mintun, E.; Ravi, N.; Mao, H.; Rolland, C.; Gustafson, L.; Xiao, T.; Whitehead, S.; Berg, A.~C.; Lo, W.-Y.; et~al. 2023.
\newblock Segment anything.
\newblock In \emph{Proceedings of the IEEE/CVF international conference on computer vision}, 4015--4026.

\bibitem[{Kirk et~al.(2024)Kirk, Mediratta, Nalmpantis, Luketina, Hambro, Grefenstette, and Raileanu}]{kirk2024understandingeffectsrlhfllm}
Kirk, R.; Mediratta, I.; Nalmpantis, C.; Luketina, J.; Hambro, E.; Grefenstette, E.; and Raileanu, R. 2024.
\newblock Understanding the Effects of RLHF on LLM Generalisation and Diversity.
\newblock arXiv:2310.06452.

\bibitem[{Li et~al.(2023)Li, Liu, Wu, Mu, Yang, Gao, Li, and Lee}]{gligen}
Li, Y.; Liu, H.; Wu, Q.; Mu, F.; Yang, J.; Gao, J.; Li, C.; and Lee, Y.~J. 2023.
\newblock Gligen: Open-set grounded text-to-image generation.
\newblock In \emph{Proceedings of the IEEE/CVF Conference on Computer Vision and Pattern Recognition}, 22511--22521.

\bibitem[{Liu et~al.(2024)Liu, Ma, Zhen, Dan, Yu, Zhao, Hu, Liu, and Fan}]{liu2024llm4genleveragingsemanticrepresentation}
Liu, M.; Ma, Y.; Zhen, Y.; Dan, J.; Yu, Y.; Zhao, Z.; Hu, Z.; Liu, B.; and Fan, C. 2024.
\newblock LLM4GEN: Leveraging Semantic Representation of LLMs for Text-to-Image Generation.
\newblock arXiv:2407.00737.

\bibitem[{Ma et~al.(2024{\natexlab{a}})Ma, Xu, Tang, Jin, Zhang, Zhao, Fan, and Hu}]{characteradapter}
Ma, Y.; Xu, W.; Tang, J.; Jin, Q.; Zhang, R.; Zhao, Z.; Fan, C.; and Hu, Z. 2024{\natexlab{a}}.
\newblock Character-Adapter: Prompt-Guided Region Control for High-Fidelity Character Customization.
\newblock arXiv:2406.16537.

\bibitem[{Ma et~al.(2024{\natexlab{b}})Ma, Xu, Zhao, Sun, Jin, Zhao, Fan, and Hu}]{Storynizor}
Ma, Y.; Xu, W.; Zhao, C.; Sun, K.; Jin, Q.; Zhao, Z.; Fan, C.; and Hu, Z. 2024{\natexlab{b}}.
\newblock Storynizor: Consistent Story Generation via Inter-Frame Synchronized and Shuffled ID Injection.
\newblock arXiv:2409.19624.

\bibitem[{Mildenhall et~al.(2021)Mildenhall, Srinivasan, Tancik, Barron, Ramamoorthi, and Ng}]{mildenhall2021nerf}
Mildenhall, B.; Srinivasan, P.~P.; Tancik, M.; Barron, J.~T.; Ramamoorthi, R.; and Ng, R. 2021.
\newblock Nerf: Representing scenes as neural radiance fields for view synthesis.
\newblock \emph{Communications of the ACM}, 65(1): 99--106.

\bibitem[{Mo et~al.(2024)Mo, Mu, Lin, Liu, Guan, Li, and Zhou}]{freecontrol}
Mo, S.; Mu, F.; Lin, K.~H.; Liu, Y.; Guan, B.; Li, Y.; and Zhou, B. 2024.
\newblock Freecontrol: Training-free spatial control of any text-to-image diffusion model with any condition.
\newblock In \emph{Proceedings of the IEEE/CVF Conference on Computer Vision and Pattern Recognition}, 7465--7475.

\bibitem[{Mou et~al.(2024)Mou, Wang, Xie, Wu, Zhang, Qi, and Shan}]{t2i}
Mou, C.; Wang, X.; Xie, L.; Wu, Y.; Zhang, J.; Qi, Z.; and Shan, Y. 2024.
\newblock T2i-adapter: Learning adapters to dig out more controllable ability for text-to-image diffusion models.
\newblock In \emph{Proceedings of the AAAI Conference on Artificial Intelligence}, volume~38, 4296--4304.

\bibitem[{Nichol et~al.(2021)Nichol, Dhariwal, Ramesh, Shyam, Mishkin, McGrew, Sutskever, and Chen}]{glide}
Nichol, A.; Dhariwal, P.; Ramesh, A.; Shyam, P.; Mishkin, P.; McGrew, B.; Sutskever, I.; and Chen, M. 2021.
\newblock Glide: Towards photorealistic image generation and editing with text-guided diffusion models.
\newblock \emph{arXiv preprint arXiv:2112.10741}.

\bibitem[{OpenAI(2024)}]{gpt-4}
OpenAI. 2024.
\newblock GPT-4 Technical Report.
\newblock arXiv:2303.08774.

\bibitem[{Ouyang et~al.(2022)Ouyang, Wu, Jiang, Almeida, Wainwright, Mishkin, Zhang, Agarwal, Slama, Ray et~al.}]{training}
Ouyang, L.; Wu, J.; Jiang, X.; Almeida, D.; Wainwright, C.; Mishkin, P.; Zhang, C.; Agarwal, S.; Slama, K.; Ray, A.; et~al. 2022.
\newblock Training language models to follow instructions with human feedback.
\newblock \emph{Advances in neural information processing systems}, 35: 27730--27744.

\bibitem[{Phung, Ge, and Huang(2024)}]{attention-refocusing}
Phung, Q.; Ge, S.; and Huang, J.-B. 2024.
\newblock Grounded text-to-image synthesis with attention refocusing.
\newblock In \emph{Proceedings of the IEEE/CVF Conference on Computer Vision and Pattern Recognition}, 7932--7942.

\bibitem[{Ramesh et~al.(2022)Ramesh, Dhariwal, Nichol, Chu, and Chen}]{DALL-E2}
Ramesh, A.; Dhariwal, P.; Nichol, A.; Chu, C.; and Chen, M. 2022.
\newblock Hierarchical text-conditional image generation with clip latents.
\newblock \emph{arXiv preprint arXiv:2204.06125}, 1(2): 3.

\bibitem[{Ramesh et~al.(2021)Ramesh, Pavlov, Goh, Gray, Voss, Radford, Chen, and Sutskever}]{DALL-E}
Ramesh, A.; Pavlov, M.; Goh, G.; Gray, S.; Voss, C.; Radford, A.; Chen, M.; and Sutskever, I. 2021.
\newblock Zero-Shot Text-to-Image Generation.
\newblock \emph{International Conference on Machine Learning,International Conference on Machine Learning}.

\bibitem[{Reed et~al.(2016)Reed, Akata, Yan, Logeswaran, Schiele, and Lee}]{gan-cls}
Reed, S.; Akata, Z.; Yan, X.; Logeswaran, L.; Schiele, B.; and Lee, H. 2016.
\newblock Generative adversarial text to image synthesis.
\newblock In \emph{International conference on machine learning}, 1060--1069. PMLR.

\bibitem[{Ren et~al.(2023)Ren, Lattas, Gecer, Deng, Ma, and Yang}]{ren2023pbidr}
Ren, X.; Lattas, A.; Gecer, B.; Deng, J.; Ma, C.; and Yang, X. 2023.
\newblock Facial Geometric Detail Recovery via Implicit Representation.
\newblock In \emph{2023 IEEE 17th International Conference on Automatic Face and Gesture Recognition (FG)}.

\bibitem[{Rombach et~al.(2022)Rombach, Blattmann, Lorenz, Esser, and Ommer}]{StableDiffusion}
Rombach, R.; Blattmann, A.; Lorenz, D.; Esser, P.; and Ommer, B. 2022.
\newblock High-Resolution Image Synthesis with Latent Diffusion Models.
\newblock In \emph{2022 IEEE/CVF Conference on Computer Vision and Pattern Recognition (CVPR)}.

\bibitem[{Sabini and Rusak(2018)}]{sabini2018painting}
Sabini, M.; and Rusak, G. 2018.
\newblock Painting outside the box: Image outpainting with gans.
\newblock \emph{arXiv preprint arXiv:1808.08483}.

\bibitem[{Saharia et~al.(2022)Saharia, Chan, Saxena, Li, Whang, Denton, Ghasemipour, Gontijo~Lopes, Karagol~Ayan, Salimans et~al.}]{Imagen}
Saharia, C.; Chan, W.; Saxena, S.; Li, L.; Whang, J.; Denton, E.~L.; Ghasemipour, K.; Gontijo~Lopes, R.; Karagol~Ayan, B.; Salimans, T.; et~al. 2022.
\newblock Photorealistic text-to-image diffusion models with deep language understanding.
\newblock \emph{Advances in neural information processing systems}, 35: 36479--36494.

\bibitem[{Shao et~al.(2024)Shao, Wang, Zhu, Xu, Song, Bi, Zhang, Zhang, Li, Wu et~al.}]{deepseekmath}
Shao, Z.; Wang, P.; Zhu, Q.; Xu, R.; Song, J.; Bi, X.; Zhang, H.; Zhang, M.; Li, Y.; Wu, Y.; et~al. 2024.
\newblock Deepseekmath: Pushing the limits of mathematical reasoning in open language models.
\newblock \emph{arXiv preprint arXiv:2402.03300}.

\bibitem[{Shen et~al.(2023)Shen, Song, Tan, Li, Lu, and Zhuang}]{hugginggpt}
Shen, Y.; Song, K.; Tan, X.; Li, D.; Lu, W.; and Zhuang, Y. 2023.
\newblock Hugginggpt: Solving ai tasks with chatgpt and its friends in hugging face.
\newblock \emph{Advances in Neural Information Processing Systems}, 36: 38154--38180.

\bibitem[{Tao et~al.(2022)Tao, Tang, Wu, Jing, Bao, and Xu}]{df-gan}
Tao, M.; Tang, H.; Wu, F.; Jing, X.-Y.; Bao, B.-K.; and Xu, C. 2022.
\newblock Df-gan: A simple and effective baseline for text-to-image synthesis.
\newblock In \emph{Proceedings of the IEEE/CVF conference on computer vision and pattern recognition}, 16515--16525.

\bibitem[{Team(2025)}]{hunyuan3d22025tencent}
Team, T.~H. 2025.
\newblock Hunyuan3D 2.0: Scaling Diffusion Models for High Resolution Textured 3D Assets Generation.
\newblock arXiv:2501.12202.

\bibitem[{Touvron et~al.(2023)Touvron, Lavril, Izacard, Martinet, Lachaux, Lacroix, Rozi{\`e}re, Goyal, Hambro, Azhar et~al.}]{llama}
Touvron, H.; Lavril, T.; Izacard, G.; Martinet, X.; Lachaux, M.-A.; Lacroix, T.; Rozi{\`e}re, B.; Goyal, N.; Hambro, E.; Azhar, F.; et~al. 2023.
\newblock Llama: Open and efficient foundation language models.
\newblock \emph{arXiv preprint arXiv:2302.13971}.

\bibitem[{Wei et~al.(2022)Wei, Tay, Bommasani, Raffel, Zoph, Borgeaud, Yogatama, Bosma, Zhou, Metzler et~al.}]{wei2022emergent}
Wei, J.; Tay, Y.; Bommasani, R.; Raffel, C.; Zoph, B.; Borgeaud, S.; Yogatama, D.; Bosma, M.; Zhou, D.; Metzler, D.; et~al. 2022.
\newblock Emergent abilities of large language models.
\newblock \emph{arXiv preprint arXiv:2206.07682}.

\bibitem[{Wu et~al.(2024{\natexlab{a}})Wu, Huang, Ji, Li, Cai, Kuang, Liu, Sun, and Ji}]{TraDiffusion}
Wu, M.; Huang, O.; Ji, J.; Li, J.; Cai, X.; Kuang, H.; Liu, J.; Sun, X.; and Ji, R. 2024{\natexlab{a}}.
\newblock TraDiffusion: Trajectory-Based Training-Free Image Generation.
\newblock arXiv:2408.09739.

\bibitem[{Wu et~al.(2024{\natexlab{b}})Wu, Bansal, Zhang, Wu, Li, Zhu, Jiang, Zhang, Zhang, Liu et~al.}]{wu2024autogen}
Wu, Q.; Bansal, G.; Zhang, J.; Wu, Y.; Li, B.; Zhu, E.; Jiang, L.; Zhang, X.; Zhang, S.; Liu, J.; et~al. 2024{\natexlab{b}}.
\newblock Autogen: Enabling next-gen LLM applications via multi-agent conversations.
\newblock In \emph{First Conference on Language Modeling}.

\bibitem[{Xia et~al.(2023)Xia, Zhang, Wang, Wang, Wu, Tian, Yang, and Van~Gool}]{xia2023diffir}
Xia, B.; Zhang, Y.; Wang, S.; Wang, Y.; Wu, X.; Tian, Y.; Yang, W.; and Van~Gool, L. 2023.
\newblock Diffir: Efficient diffusion model for image restoration.
\newblock In \emph{Proceedings of the IEEE/CVF International Conference on Computer Vision}, 13095--13105.

\bibitem[{Xie et~al.(2023)Xie, Li, Huang, Liu, Zhang, Zheng, and Shou}]{boxdiff}
Xie, J.; Li, Y.; Huang, Y.; Liu, H.; Zhang, W.; Zheng, Y.; and Shou, M.~Z. 2023.
\newblock Boxdiff: Text-to-image synthesis with training-free box-constrained diffusion.
\newblock In \emph{Proceedings of the IEEE/CVF International Conference on Computer Vision}, 7452--7461.

\bibitem[{Xu et~al.(2018)Xu, Zhang, Huang, Zhang, Gan, Huang, and He}]{attngan}
Xu, T.; Zhang, P.; Huang, Q.; Zhang, H.; Gan, Z.; Huang, X.; and He, X. 2018.
\newblock Attngan: Fine-grained text to image generation with attentional generative adversarial networks.
\newblock In \emph{Proceedings of the IEEE conference on computer vision and pattern recognition}, 1316--1324.

\bibitem[{Xu et~al.(2025)Xu, Yang, Wang, Hu, Wu, Wang, Luo, Zhang, Hu, and Zhang}]{xu2025comfyui}
Xu, Z.; Yang, X.; Wang, Y.; Hu, Q.; Wu, Z.; Wang, L.; Luo, W.; Zhang, K.; Hu, B.; and Zhang, M. 2025.
\newblock ComfyUI-Copilot: An Intelligent Assistant for Automated Workflow Development.
\newblock \emph{arXiv preprint arXiv:2506.05010}.

\bibitem[{Xue et~al.(2024)Xue, Lu, Huang, Wang, Ouyang, and Bai}]{comfybench}
Xue, X.; Lu, Z.; Huang, D.; Wang, Z.; Ouyang, W.; and Bai, L. 2024.
\newblock ComfyBench: Benchmarking LLM-based Agents in ComfyUI for Autonomously Designing Collaborative AI Systems.
\newblock \emph{arXiv preprint arXiv:2409.01392}.

\bibitem[{Yang et~al.(2024)Yang, Yang, Zhang, Hui, Zheng, Yu, Li, Liu, Huang, Wei, Lin, Yang, Tu, Zhang, Yang, Yang, Zhou, Lin, Dang, Lu, Bao, Yang, Yu, Li, Xue, Zhang, Zhu, Men, Lin, Li, Xia, Ren, Ren, Fan, Su, Zhang, Wan, Liu, Cui, Zhang, and Qiu}]{qwen2.5}
Yang, A.; Yang, B.; Zhang, B.; Hui, B.; Zheng, B.; Yu, B.; Li, C.; Liu, D.; Huang, F.; Wei, H.; Lin, H.; Yang, J.; Tu, J.; Zhang, J.; Yang, J.; Yang, J.; Zhou, J.; Lin, J.; Dang, K.; Lu, K.; Bao, K.; Yang, K.; Yu, L.; Li, M.; Xue, M.; Zhang, P.; Zhu, Q.; Men, R.; Lin, R.; Li, T.; Xia, T.; Ren, X.; Ren, X.; Fan, Y.; Su, Y.; Zhang, Y.; Wan, Y.; Liu, Y.; Cui, Z.; Zhang, Z.; and Qiu, Z. 2024.
\newblock Qwen2.5 Technical Report.
\newblock \emph{arXiv preprint arXiv:2412.15115}.

\bibitem[{Ye et~al.(2023)Ye, Zhang, Liu, Han, and Yang}]{ip-adapter}
Ye, H.; Zhang, J.; Liu, S.; Han, X.; and Yang, W. 2023.
\newblock Ip-adapter: Text compatible image prompt adapter for text-to-image diffusion models.
\newblock \emph{arXiv preprint arXiv:2308.06721}.

\bibitem[{Yu et~al.(2022)Yu, Xu, Koh, Luong, Baid, Wang, Vasudevan, Ku, Yang, Ayan et~al.}]{scaling}
Yu, J.; Xu, Y.; Koh, J.~Y.; Luong, T.; Baid, G.; Wang, Z.; Vasudevan, V.; Ku, A.; Yang, Y.; Ayan, B.~K.; et~al. 2022.
\newblock Scaling autoregressive models for content-rich text-to-image generation.
\newblock \emph{arXiv preprint arXiv:2206.10789}, 2(3): 5.

\bibitem[{Yuan et~al.(2024)Yuan, Liu, Cao, Sun, Jia, Chen, Li, Lin, Yuan, He et~al.}]{mora}
Yuan, Z.; Liu, Y.; Cao, Y.; Sun, W.; Jia, H.; Chen, R.; Li, Z.; Lin, B.; Yuan, L.; He, L.; et~al. 2024.
\newblock Mora: Enabling generalist video generation via a multi-agent framework.
\newblock \emph{arXiv preprint arXiv:2403.13248}.

\bibitem[{Zeng et~al.(2022)Zeng, Liu, Du, Wang, Lai, Ding, Yang, Xu, Zheng, Xia et~al.}]{glm}
Zeng, A.; Liu, X.; Du, Z.; Wang, Z.; Lai, H.; Ding, M.; Yang, Z.; Xu, Y.; Zheng, W.; Xia, X.; et~al. 2022.
\newblock Glm-130b: An open bilingual pre-trained model.
\newblock \emph{arXiv preprint arXiv:2210.02414}.

\bibitem[{Zeng et~al.(2021)Zeng, Lin, Lu, and Patel}]{zeng2021cr}
Zeng, Y.; Lin, Z.; Lu, H.; and Patel, V.~M. 2021.
\newblock Cr-fill: Generative image inpainting with auxiliary contextual reconstruction.
\newblock In \emph{Proceedings of the IEEE/CVF international conference on computer vision}, 14164--14173.

\bibitem[{Zhang et~al.(2021)Zhang, Koh, Baldridge, Lee, and Yang}]{zhang2021cross}
Zhang, H.; Koh, J.~Y.; Baldridge, J.; Lee, H.; and Yang, Y. 2021.
\newblock Cross-modal contrastive learning for text-to-image generation.
\newblock In \emph{Proceedings of the IEEE/CVF conference on computer vision and pattern recognition}, 833--842.

\bibitem[{Zhang et~al.(2017)Zhang, Xu, Li, Zhang, Wang, Huang, and Metaxas}]{stackgan}
Zhang, H.; Xu, T.; Li, H.; Zhang, S.; Wang, X.; Huang, X.; and Metaxas, D.~N. 2017.
\newblock Stackgan: Text to photo-realistic image synthesis with stacked generative adversarial networks.
\newblock In \emph{Proceedings of the IEEE international conference on computer vision}, 5907--5915.

\bibitem[{Zhang et~al.(2018)Zhang, Xu, Li, Zhang, Wang, Huang, and Metaxas}]{stackgan++}
Zhang, H.; Xu, T.; Li, H.; Zhang, S.; Wang, X.; Huang, X.; and Metaxas, D.~N. 2018.
\newblock Stackgan++: Realistic image synthesis with stacked generative adversarial networks.
\newblock \emph{IEEE transactions on pattern analysis and machine intelligence}, 41(8): 1947--1962.

\bibitem[{Zhang, Rao, and Agrawala(2023)}]{controlnet}
Zhang, L.; Rao, A.; and Agrawala, M. 2023.
\newblock Adding conditional control to text-to-image diffusion models.
\newblock In \emph{Proceedings of the IEEE/CVF International Conference on Computer Vision}, 3836--3847.

\bibitem[{Zhang et~al.(2022)Zhang, Roller, Goyal, Artetxe, Chen, Chen, Dewan, Diab, Li, Lin et~al.}]{opt}
Zhang, S.; Roller, S.; Goyal, N.; Artetxe, M.; Chen, M.; Chen, S.; Dewan, C.; Diab, M.; Li, X.; Lin, X.~V.; et~al. 2022.
\newblock Opt: Open pre-trained transformer language models.
\newblock \emph{arXiv preprint arXiv:2205.01068}.

\bibitem[{Zhao et~al.(2023)Zhao, Li, Jin, and Zhou}]{loco}
Zhao, P.; Li, H.; Jin, R.; and Zhou, S.~K. 2023.
\newblock Loco: Locally constrained training-free layout-to-image synthesis.
\newblock \emph{arXiv preprint arXiv:2311.12342}.

\bibitem[{Zhu et~al.(2019)Zhu, Pan, Chen, and Yang}]{dm-gan}
Zhu, M.; Pan, P.; Chen, W.; and Yang, Y. 2019.
\newblock Dm-gan: Dynamic memory generative adversarial networks for text-to-image synthesis.
\newblock In \emph{Proceedings of the IEEE/CVF conference on computer vision and pattern recognition}, 5802--5810.

\end{thebibliography}
\clearpage
\setcounter{page}{1}

\section{Dataset Details}
We divide the construction process of FlowDataset into six stages: Data Crawling, Data Cleaning, Semantic Enhancement, Category Summary, Data Classification, and Dataset Partitioning.

\noindent\textbf{Data Crawling.} Our data mainly comes from ComfyUI community websites like OpenArt~\footnote{\url{https://openart.ai/workflows/home}}, LibLib~\footnote{\url{https://www.liblib.art}}, ComfyWorkflows~\footnote{\url{https://comfyworkflows.com/}}, and Civital~\footnote{\url{https://civitai.com}}. Each metadata entry is created using web crawling technology and includes a parent title, title, description, tags, and a JSON-formatted ComfyUI workflow.

\noindent\textbf{Data Cleaning.} In this stage, we focus on cleaning the JSON-formatted ComfyUI workflows. This process is broken down into five steps: format validation, special node processing, removal of redundant information, connected graph assessment, and validation of workflow data. 

During the format validation process, we filter out junk data based on the ComfyUI workflow JSON schema. 

We handle certain special nodes, such as the ComfyUI node ``Anything Everywhere", which can implicitly link to any node with a matching type, and nodes with a ``bypass" mode that are skipped during the ComfyUI workflow process. Then, we remove these ambiguous nodes and restore the workflow processing by establishing explicit links. 

During the redundant information removal phase, we focus on deleting nodes that lack meaningful information, such as ``Note" and ``Reroute". 

Since workflows can be abstracted into an undirected graph structure, we filter out data that does not form a connected graph, as such data typically contains multiple distinct workflows within a single ComfyUI workflow JSON file, significantly increasing training complexity. 

Finally, based on our node base, we conduct a preliminary validation of workflow execution. This primarily checks the information of the workflows, such as whether each node's required inputs are met and whether the parameter types passed between nodes are consistent.

\begin{figure}[t]
  \centering
  \includegraphics[width=0.95\linewidth]{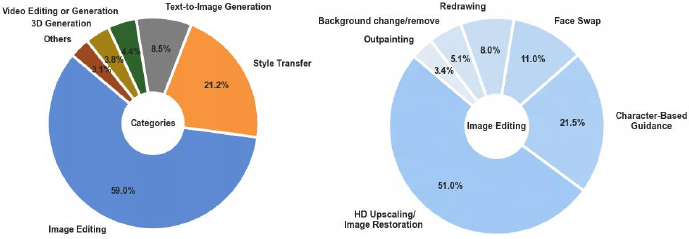}
  \caption{\textbf{Illustration of categories included in FlowBench.} The left figure represents the proportion of the six categories in Flowbench, while the right figure represents subcategories.}
  \label{fig:bench}
  \vspace{-2mm}
\end{figure}

\begin{figure}[t]
  \centering
  \includegraphics[width=0.9\linewidth]{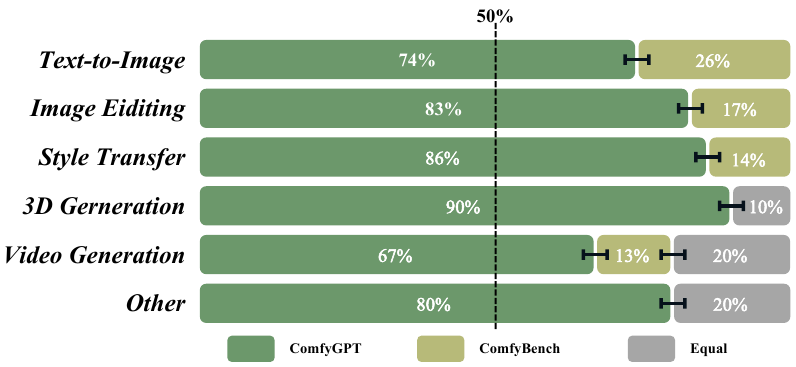}
  \caption{\textbf{User Study of ComfyGPT and ComfyBench.}}
  \label{fig:user_study}
\end{figure}

\noindent\textbf{Semantic Enhancement.} The workflow description in the original crawled data is quite noisy, which significantly increases the difficulty of training. Thanks to the powerful language understanding capabilities of LLM, we use ChatGPT-4o-mini~\cite{gpt-4} to analyze and understand the chaotic information, then polish it into a concise functional description. As shown in Table~\ref{tab:polish}, we present the prompt design.

\noindent\textbf{Category Summary} We further categorize our data, as shown in the Category Summary part of Tab.~\ref{tab:category}, we use ChatGPT-4o-mini to summarize and categorize the data based on polished descriptions. Then, we manually summarize six major categories, including: Text-to-Image Generation, Image Editing, Style Transfer, 3D Generation, Video Editing or Generation, and others. Besides, we further divided the image editing category into six subcategories, including: HD Upscaling/Image Restoration, Redrawing, Outpainting, Character-Based Guidance, Face Swap, and Background Change/Remove. 

\paragraph{Data Classification.} Based on previous categories and data-polished descriptions, we use ChatGPT-4omini to classify the data. As shown in the Category Partitioning part of Tab.~\ref{tab:category}, we present the prompt design.

\noindent\textbf{Dataset Partitioning.} After previous stages, we construct the FlowDataset consisting of a total of 13,571 entries, each comprising a polished description, a JSON-formatted workflow, and the category it belongs to. We then conduct execution testing on the flow dataset. Specifically, we test the data within each of the six major categories we have defined, ensuring that at least 70\% of the data in each category can be executed successfully by the ComfyUI server. Based on these successfully executed examples, as shown in Fig.~\ref{fig:bench}, we sample 1,000 data points proportionally from each category while also considering the distribution of workflow lengths. We then divide these 1,000 data points into FlowBench, while the remaining 12,571 serve as the training set for the FlowDataset. 


\begin{table}[t]  
    \centering
    \scalebox{0.8}{\begin{tabular}{c|ccccc}
    \toprule
  \textbf{k =} & \textbf{0}& \textbf{1}&\textbf{3}&\textbf{5}&\textbf{7}\\
    \midrule
   \textbf{PA(\%)$\uparrow$}&83.5 &84.6&85.3&\textbf{85.6}&85.4\\
    \bottomrule
    
    \end{tabular}}
    
     \vspace{-2mm}
     \caption{\textbf{Ablation Study on the Top-K Hyperparameter in RefineAgent.}}
    \label{table:abl_k}
    \vspace{-0.2cm}
\end{table}
\section{Implement Details}
\subsection{ComfyGPT}
\noindent\textbf{FlowAgent.} We utilize the original checkpoint of Qwen2.5-14B~\cite{qwen2.5} as the backbone for FlowAgent in the final version of ComfyGPT. During the SFT stage, we train the FlowAgent using FlowDataset, a total of 12,571 data entries. The training is conducted on 4 80GB NVIDIA A100 GPUs over three epochs, with a learning rate of 5e-5 and a batch size of 1. In the RL stage, we utilize 8 80GB NVIDIA A100 GPUs to train for a total of 300 steps using the same training set as in SFT, with a learning rate of 1e-6 and a batch size of 8. The number of group computations $G$ is set to 5, the clipping coefficient $\varepsilon$ to 0.2, and the KL penalty coefficient $\beta$ to 0.001. During inference, we utilize a 80GB NVIDIA A100 GPU, setting the maximum tokens to 8192, top-p to 0.7, and temperature to 0.95.

\noindent\textbf{The workflow generation process of ComfyGPT.} As shown in Tab.~\ref{tab:over}, FlowAgent can accurately generate workflow diagrams based on user instructions. However, some nodes may be incorrect due to outdated training data or fictitious entries from FlowAgent. To address this challenge, as shown in Tab.~\ref{tab:over}, RefineAgent first identifies the incorrect nodes and retrieves the top k=5 similar candidate nodes based on their names. Finally, as demonstrated in Tab.~\ref{tab:refineagent}, RefineAgent combines the user instruction, workflow diagram, incorrect node name, and candidate node information to determine the most appropriate node, leveraging the advanced language understanding capabilities of the large language model (LLM).
\subsection{Experiment details of baselines}
\noindent\textbf{The few-shot learning.} As shown in Tab.~\ref{tab:one-shot}, based on six categories, we randomly selected one example from the training set of FlowDataset for each category to provide to the LLM for few-shot learning in each inference. to generate new workflow diagrams according to the instructions in FlowBench. The LLM then generates new workflow diagrams based on the instructions in FlowBench.

\noindent\textbf{Implement Details of baselines.} For the closed-source baselines (ChatGPT-4-32~\cite{gpt-4}, ChatGPT-4o~\cite{gpt-4}, Claude-3-5-sonnet~\cite{Claude}, and Claude-3-7-sonnet~\cite{Claude3-7}), we evaluate their performance using a few-shot learning approach. For ComfyAgent~\cite{comfybench}, we follow the settings described in its original methodology. For the Multi-Agent (debate)~\cite{muti-agent-debate} method, we also use a few-shot learning approach and adopt its default hyperparameter settings. For LLaMA-13B~\cite{llama}, Baichuan2-13B-Base~\cite{baichuan2}, and ChatGLM3-6B~\cite{glm2024chatglm}, we evaluate their performance using the ComfyGPT method without the GRPO algorithm.

\noindent\textbf{The evaluation of Pass Instruct Alignment(PIA).} As shown in Tab.~\ref{tab:confirm}, we evaluate PIA of generated ComfyUI workflow using ChatGPT-4.0 mini.

\begin{figure}[t]
  \centering
  \includegraphics[width=0.95\linewidth]{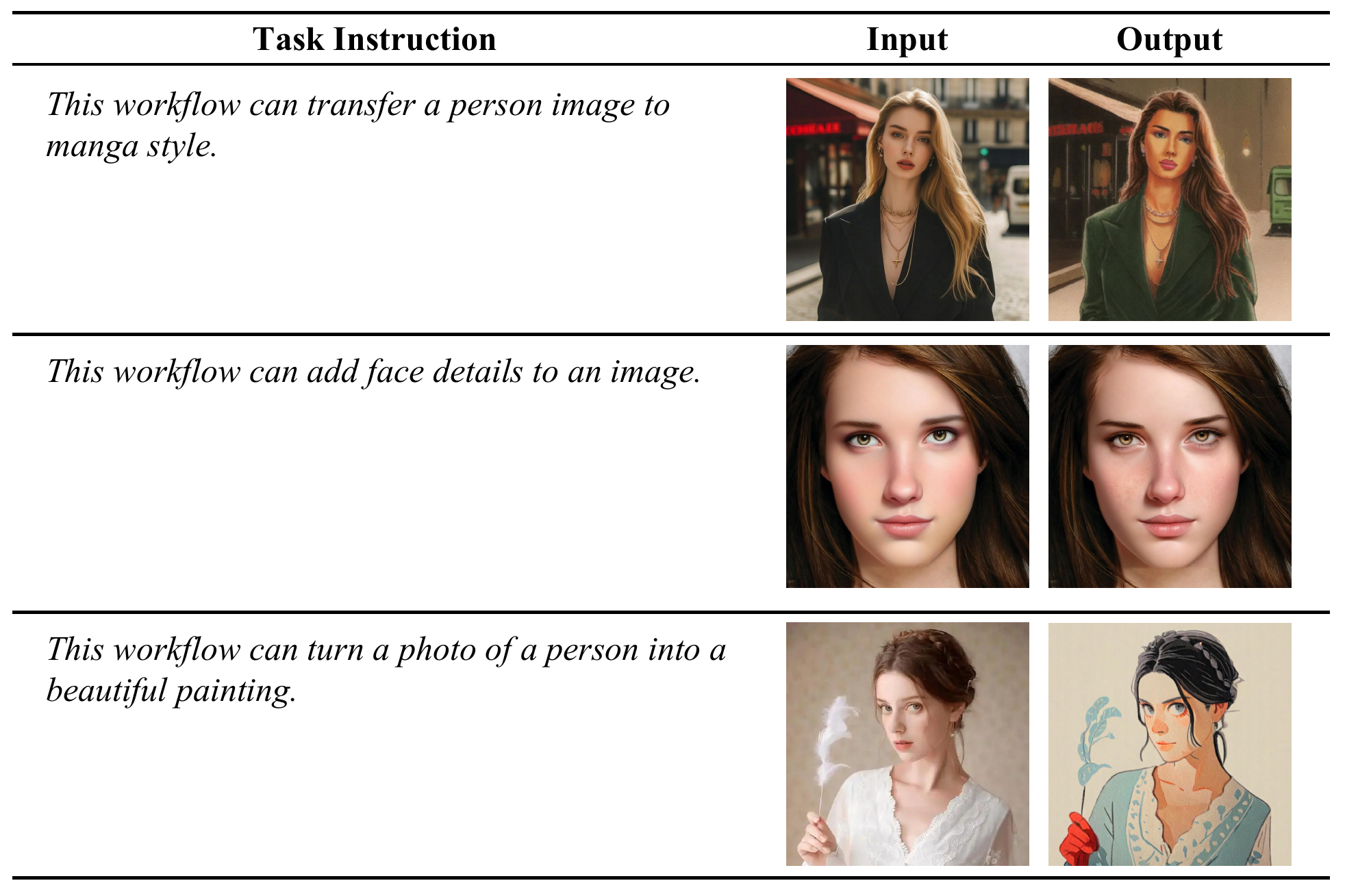}
  \caption{\textbf{More Qualitative results of ComfyGPT.}}
  \label{fig:more}
\end{figure}

\begin{figure}[t]
  \centering
  \includegraphics[width=0.95\linewidth]{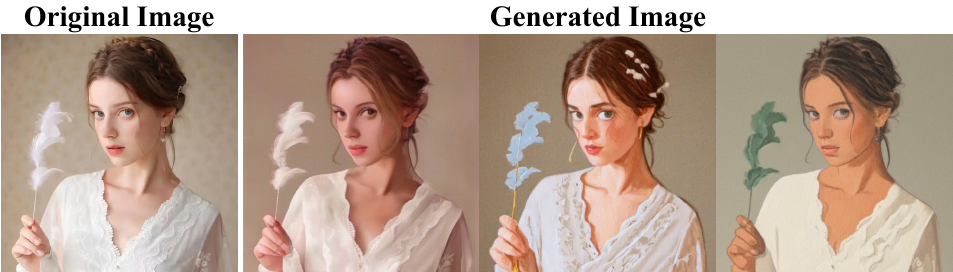}
  \caption{\textbf{Visualization of style transfer tasks under human-loop interaction.}}
  \label{fig:sty}
\end{figure}

\section{More Experiments}
\subsection{Ablations}
\noindent\textbf{Ablation study on hyperparameter $k$.} As shown in Tab~\ref{table:abl_k}, PA improves from 84.6\% to 85.6\% and then slightly drops to 85.4\% as $k$ increases from 1 to 5 to 7.

\subsection{User Study}
We conduct a user study with 20 ComfyUI users. As shown in Fig.~\ref{fig:user_study}, ComfyGPT significantly outperforms ComfyBench across multiple tasks (average over $53\%$).

\begin{figure*}[t]
  \centering
  \includegraphics[width=0.95\linewidth]{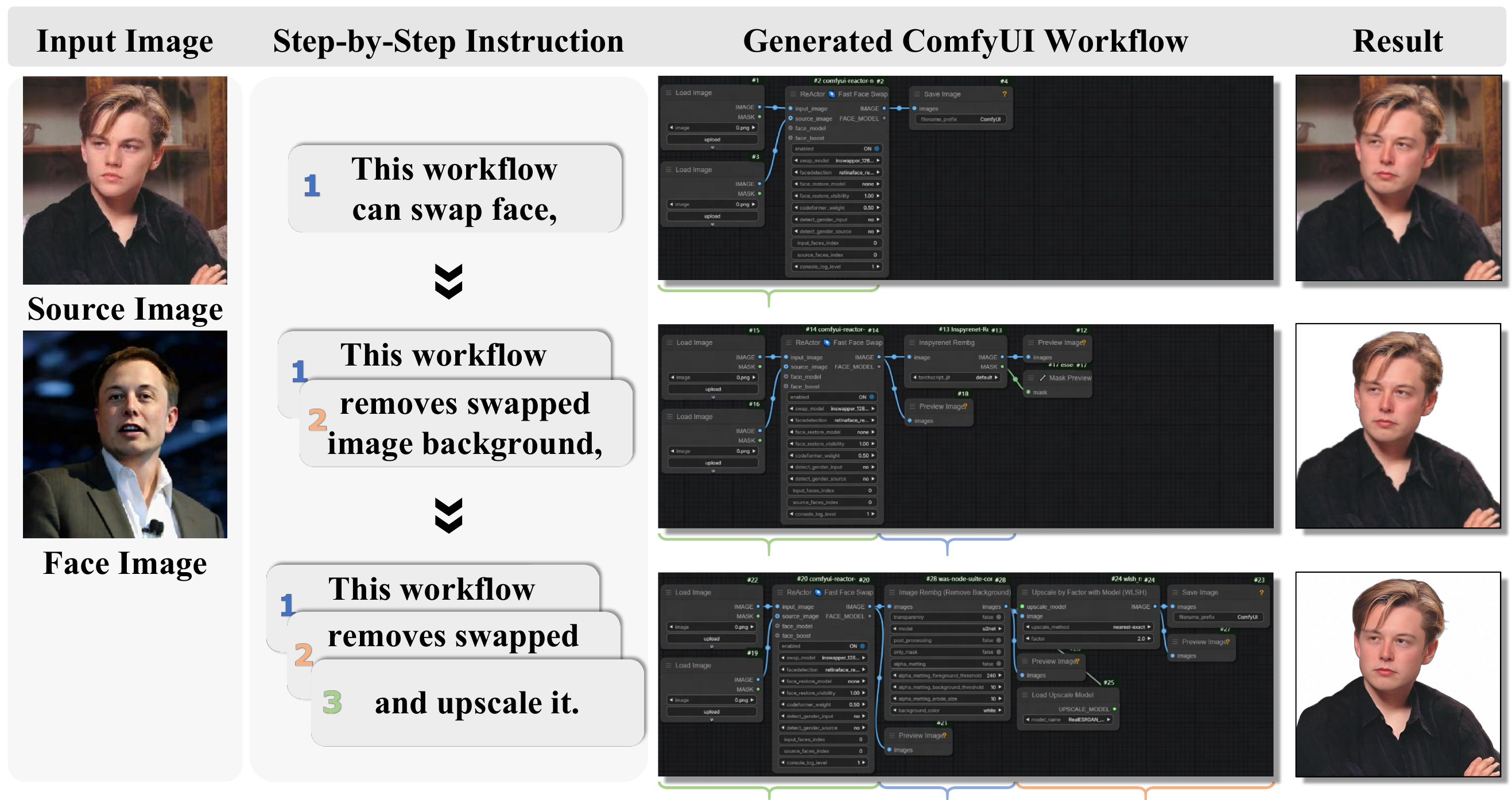}
   \caption{\textbf{Step-by-step ComfyUI workflow generation by ComfyGPT based on user instructions.} This figure illustrates the strong semantic alignment capability of ComfyGPT instructions with workflows, enabling the construction of workflows based on step-by-step instructions.}
  \label{fig:semantic}
\end{figure*}
\subsection{Qualitative Results}

\noindent\textbf{Additional Qualitative Results.} As shown in Fig.~\ref{fig:more}, we present some additional qualitative results of ComfyGPT, showcasing its diverse task capabilities.

\noindent\textbf{Task Awareness in Instructions.} As shown in Fig.~\ref{fig:semantic}, ComfyGPT possesses a robust ability to align instructions with workflows. It can clearly differentiate the functions of each component within the workflow and, as tasks within the instructions accumulate, it excels at generating appropriate and coherent workflows.

\begin{figure*}[t]
  \centering
  \includegraphics[width=0.95\linewidth]{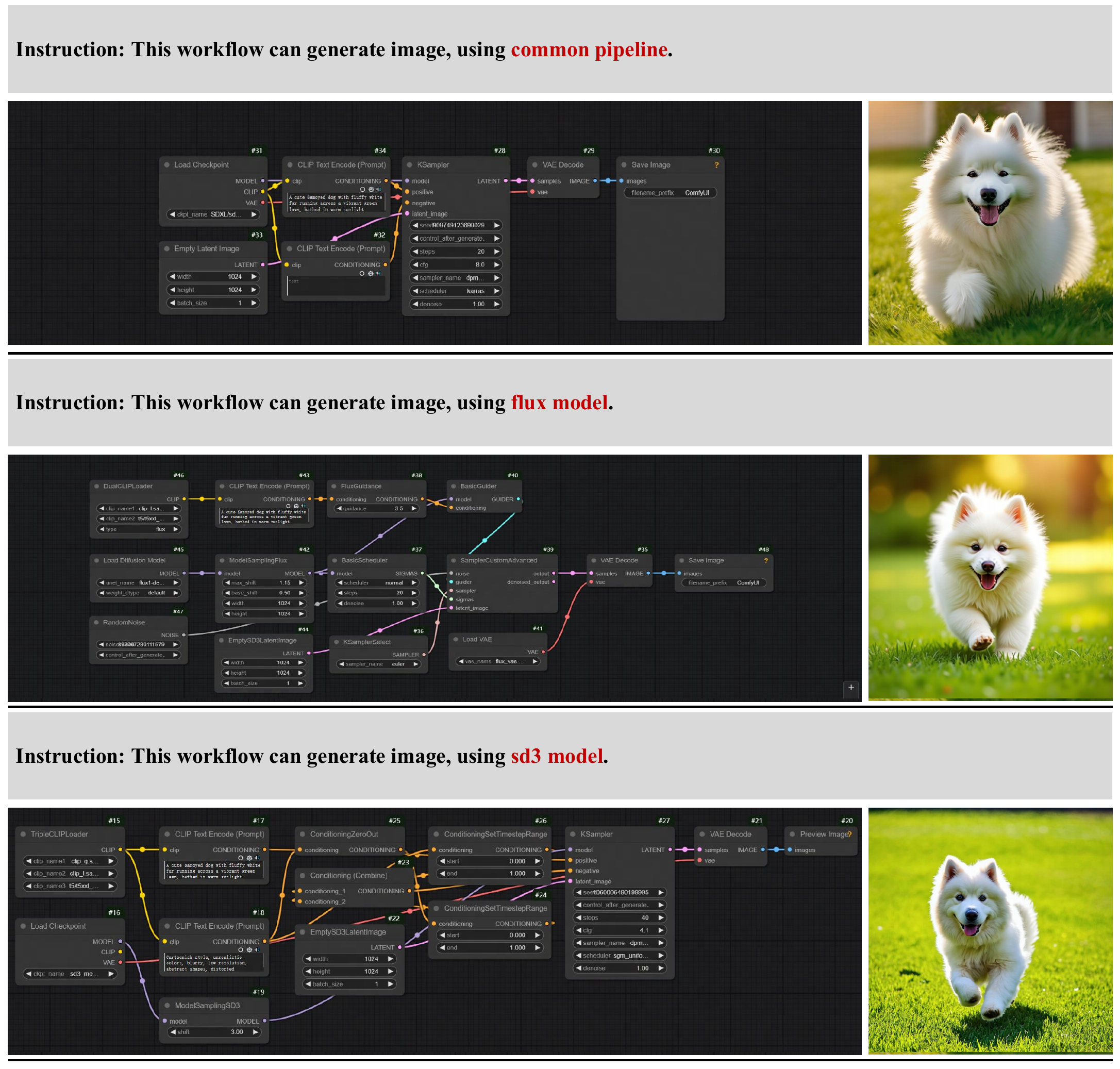}
  \caption{\textbf{Qualitative results of text-to-image workflow generation in ComfyGPT using different model keywords in instructions.}}
  \label{fig:diff_model}
\end{figure*}

\noindent\textbf{Model Awareness in Instructions.} As shown in Fig.~\ref{fig:diff_model}, ComfyGPT can automatically generate workflows that align with the corresponding model based on the keywords present in the instructions.

\section{Discussion}
\noindent\textbf{Reduction in node diversity after GRPO.} The reduction in the node diversity is a common phenomenon in RL as the model becomes more convergent and less diverse\cite{kirk2024understandingeffectsrlhfllm}. Reinforcement learning aligns the model with user preferences, leading to more deterministic and valid node generation. As shown in Tab. 4 in the main paper, FV, PA, and PIA improve, showing that GRPO helps the model generate more stable and accurate workflows, with only a minimal impact on node diversity. This indicates GRPO improves the generation performance overall. 

\noindent\textbf{Potential error accumulation across components.} We address error accumulation as follows: 1) ReformatAgent and ExecuteAgent only handle format conversion and do not introduce semantic errors; 2) FlowAgent is trained via supervised learning for accurate workflow generation; 3) RefineAgent uses RAG to correct outdated or invalid nodes. Therefore, compared with ComfyBench~\cite{comfybench}, our method shows consistently better performance (Tab. 3 in the main paper), demonstrating stronger robustness to error propagation.

\noindent\textbf{Uncertain future of ComfyUI~\footnote{\url{https://github.com/comfyanonymous/ComfyUI}}.} While future trends may evolve, ComfyUI is currently the most widely adopted platform, with \textbf{76.7k} GitHub stars and a strong ecosystem (e.g., OpenArt, Liblib). Building on it is both timely and impactful, addressing real-world needs.

\noindent\textbf{Human-in-the-loop interactions in ComfyGPT.} We indeed consider the human-in-the-loop interactions ComfyGPT is designed to automatically generate workflows executable on the ComfyUI server, while the actual execution often requires user-specific parameter tuning. For instance, in style transfer tasks (Fig.~\ref{fig:sty}), control strength depends on user preferences. Our framework allows users to freely adjust hyperparameters.

\noindent\textbf{Why is node diversity important?} Node diversity indicates the variety of generated workflows. For instance, in a face swap workflow, nodes like InstantID, Pulid, or IP-Adapter can be used. A decline in node diversity indicates reduced workflow variety.

\noindent\textbf{The significance of Nodebase.} The node database does not introduce significant complexity, and it is important for our system: 1) it is updated infrequently (once a month), and updates can be easily retrieved from the ComfyUI server. New algorithms and nodes in the open-source community can be easily updated in our node database, making ComfyGPT better access to new technology; 2) The node database functions as an information retrieval tool in the RefineAgent to help correct node generation and brings the improvement of the accuracy, as shown in Tab. 4 in the main paper. 

\section{Related Works}
\noindent\textbf{Multi-Agent frameworks.} The research landscape of multi-agent systems has expanded rapidly, fuelled by both advances in reinforcement learning and the recent surge of large language models (LLMs). On the one hand, several general-purpose frameworks have been proposed. A representative example is Multi-Agent(Debate)~\cite{muti-agent-debate}, which orchestrates multiple large language models to solve problems through iterative rounds of debate that eventually converge on a high-quality answer. AutoGen~\cite{wu2024autogen} generalizes the idea of “LLM societies” by allowing an arbitrary number of role-playing agents to converse, plan, and call external tools. Its declarative graphs specify which agents communicate, under what conditions, and with what stopping criteria, enabling complex workflows such as iterative code generation and data annotation. On the other hand, a range of domain-specific systems has been developed. In these systems, each agent is tailored to a particular task, which makes cross-domain transfer difficult. Mora~\cite{mora} and GenMac~\cite{genmac} focus on video generation with different goals and agent structures. Mora~\cite{mora} aims to reproduce a closed-source system, Sora~\footnote{\url{https://openai.com/sora/}}, focusing on minute-long text-to-video generation. In its architecture, the Prompt Selection Agent enhances textual semantics, while the Text-to-Image and Image-to-Image Agents generate the first video frame tailored to user preferences. The Image-to-Video and Video Transition Agents are responsible for producing and composing video segments. The entire process relies heavily on close user interaction. GenMac~\cite{genmac} addresses text-to-video generation for compositional prompts. Our method focuses on generating ComfyUI workflows, which differ from their task. The design and functionality of each agent in our approach are distinct.

\begin{table*}[t]
    \centering
    \renewcommand{\arraystretch}{1.2} 
    \scalebox{1}{\begin{tabular}{|p{14cm}|}
        \hline
        \textbf{Semantic Enhancement} \\ 
        \hline
        \small
        \#\textbf{Prompt} - I would like you to act as an expert in information processing. I will provide a chaotic infomation regarding an computer vision task workflow. Your task is to analyze and understand it, then summarize the core functionality. Infomation:
        \{\{\textit{Infomation}\}\} Please return the results in pure JSON format, including the following: 1.summary: A refined description of the functionality, limited to 100 words. If you cannot analyze and extract valid any concept. you should return an empty string. 
        
        Response Example:\\
        \texttt{```}json\{\texttt{"}summary\texttt{"}: \texttt{"}The workflow allows uploading photos and converting them into stylized images.\texttt{"}\} \texttt{```} \\

        \hline
    \end{tabular}}
    \caption{\textbf{The prompt design of the Semantic Enhancement.} In the prompt, we define some injectable slots, such as \{\{\textit{Infomation}\}\}, which are dynamically replaced by specific content during execution.}
    \label{tab:polish}
\end{table*}

\begin{table*}[t]
    \centering
    \renewcommand{\arraystretch}{1.2} 
    \scalebox{1}{\begin{tabular}{|p{14cm}|}
        \hline
        \textbf{Category Summary} \\ 
        \hline
        \small
        \#\textbf{Prompt} - I would like you to act as an expert in information processing. I will provide a description regarding an image processing workflow. Your task is to analyze and understand it, and summary a computer vision task category about it. Description:\{\{\textit{Description}\}\} Please return the results in JSON format, including the following: 1.belong\_category: It indicates which task category the description belongs to. Notes: 1.You should not create categories with very broad concepts. These categories should belong to a specific task in the field of computer vision. 2.You should return the result in pure JSON format without including any other information or code. 
        
        Response Example: 
              
        \texttt{```}json\{\texttt{"}belong\_category\texttt{"}: \texttt{"}text\texttt{-}to\texttt{-}image\texttt{"}\}\texttt{```}
        \\

        \hline
        \textbf{Data Classification} \\ 

        \hline
        \small
        \#\textbf{Prompt} - I would like you to act as an expert in information processing. I will provide a description regarding a computer vision task and some computer vision task categories. Your task is to analyze and understand the description and these task categories, and determine which of the task category I provided the description belongs to. Description: \{\{\textit{Description}\}\} Categoryies: \{\{\textit{Categoryies}\}\} Please return the results in JSON format, including the following: 1.belong\_category: It indicates which category the description belongs to. This category must be included in the categories I provided. If none of them match, it should be classified as other. Notes: 1.You should return the result in pure JSON format without including any other information or code. 
        
        Response Example: 
        
        \texttt{```}json\{\texttt{"}belong\_category\texttt{"}: \texttt{"}Text-to-image generation\texttt{"}\}\texttt{```}\\
        \hline
    \end{tabular}}
    \caption{\textbf{The prompt design of Category Summary and Data Classification.} In the prompt, we define some injectable slots, such as \{\{\textit{Description}\}\} and \{\{\textit{Categoryies}\}\}, which are dynamically replaced by specific content during execution.}
    \label{tab:category}
\end{table*}

\begin{table*}[t]
    \centering
    \renewcommand{\arraystretch}{1.2} 
    \scalebox{1}{\begin{tabular}{|p{14cm}|}
        \hline
        \textbf{FlowAgent} \\ 
        \hline
        \small
        \#1 \textbf{Instruction} - The workflow facilitates the generation of consistent style images by allowing users to upload a character profile picture and a style reference image. It offers customizable workflows and Lora for ensuring image consistency, along with access to AI painting video tutorials. Users can enhance images using high-definition repair if needed.\\
        \small
        \#2 \textbf{Output} - [[\texttt{"}ImageResizeKJ\_0\texttt{"}, \texttt{"}IMAGE\texttt{"}, \texttt{"}easy imageSize\_0\texttt{"}, \texttt{"}image\texttt{"}], [\texttt{"}UNETLoader\_0\texttt{"}, \texttt{"}MODEL\texttt{"}, \texttt{"}LoraLoader\_0\texttt{"}, \texttt{"}model\texttt{"}], [\texttt{"}DualCLIPLoader\_0\texttt{"}, \texttt{"}CLIP\texttt{"}, \texttt{"}LoraLoader\_0\texttt{"}, \texttt{"}clip\texttt{"}], [\texttt{"}RandomNoise\_0\texttt{"}, \texttt{"}NOISE\texttt{"}, \texttt{"}SamplerCustomAdvanced\_0\texttt{"}, \texttt{"}noise\texttt{"}], [\texttt{"}BasicGuider\_0\texttt{"}, \texttt{"}GUIDER\texttt{"}, \texttt{"}SamplerCustomAdvanced\_0\texttt{"}, \texttt{"}guider\texttt{"}], [\texttt{"}LoraLoader\_0\texttt{"}, \texttt{"}CLIP\texttt{"}, \texttt{"}CLIPTextEncodeFlux\_0\texttt{"}, \texttt{"}clip\texttt{"}], [\texttt{"}KSamplerSelect\_0\texttt{"}, \texttt{"}SAMPLER\texttt{"}, \texttt{"}SamplerCustomAdvanced\_0\texttt{"}, \texttt{"}sampler\texttt{"}], [\texttt{"}BasicScheduler\_0\texttt{"}, \texttt{"}SIGMAS\texttt{"}, \texttt{"}SamplerCustomAdvanced\_0\texttt{"}, \texttt{"}sigmas\texttt{"}], [\texttt{"}SamplerCustomAdvanced\_0\texttt{"}, \texttt{"}output\texttt{"}, \texttt{"}VAEDecode\_0\texttt{"}, \texttt{"}samples\texttt{"}], [\texttt{"}VAELoader\_0\texttt{"}, \texttt{"}VAE\texttt{"}, \texttt{"}VAEDecode\_0\texttt{"}, \texttt{"}vae\texttt{"}], [\texttt{"}ModelSamplingFlux\_0\texttt{"}, \texttt{"}MODEL\texttt{"}, \texttt{"}BasicGuider\_0\texttt{"}, \texttt{"}model\texttt{"}], [\texttt{"}ModelSamplingFlux\_0\texttt{"}, \texttt{"}MODEL\texttt{"}, \texttt{"}BasicScheduler\_0\texttt{"}, \texttt{"}model\texttt{"}], [\texttt{"}easy imageSize\_1\texttt{"}, \texttt{"}width\_int\texttt{"}, \texttt{"}ModelSamplingFlux\_0\texttt{"}, \texttt{"}width\texttt{"}], [\texttt{"}easy imageSize\_1\texttt{"}, \texttt{"}height\_int\texttt{"}, \texttt{"}ModelSamplingFlux\_0\texttt{"}, \texttt{"}height\texttt{"}], [\texttt{"}CLIPTextEncodeFlux\_0\texttt{"}, \texttt{"}CONDITIONING\texttt{"}, \texttt{"}ConditioningZeroOut\_0\texttt{"}, \texttt{"}conditioning\texttt{"}], [\texttt{"}ReplaceString\_0\texttt{"}, \texttt{"}STRING\texttt{"}, \texttt{"}CLIPTextEncodeFlux\_0\texttt{"}, \texttt{"}clip\_l\texttt{"}], [\texttt{"}ReplaceString\_0\texttt{"}, \texttt{"}STRING\texttt{"}, \texttt{"}CLIPTextEncodeFlux\_0\texttt{"}, \texttt{"}t5xxl\texttt{"}], [\texttt{"}Text Multiline\_0\texttt{"}, \texttt{"}STRING\texttt{"}, \texttt{"}ReplaceString\_0\texttt{"}, \texttt{"}String\texttt{"}], [\texttt{"}Joy\_caption\_load\_0\texttt{"}, \texttt{"}JoyPipeline\texttt{"}, \texttt{"}Joy\_caption\_0\texttt{"}, \texttt{"}joy\_pipeline\texttt{"}], [\texttt{"}InpaintModelConditioning\_0\texttt{"}, \texttt{"}positive\texttt{"}, \texttt{"}BasicGuider\_0\texttt{"}, \texttt{"}conditioning\texttt{"}], [\texttt{"}ConditioningZeroOut\_0\texttt{"}, \texttt{"}CONDITIONING\texttt{"}, \texttt{"}InpaintModelConditioning\_0\texttt{"}, \texttt{"}negative\texttt{"}], [\texttt{"}VAELoader\_0\texttt{"}, \texttt{"}VAE\texttt{"}, \texttt{"}InpaintModelConditioning\_0\texttt{"}, \texttt{"}vae\texttt{"}], [\texttt{"}InpaintModelConditioning\_0\texttt{"}, \texttt{"}latent\texttt{"}, \texttt{"}SamplerCustomAdvanced\_0\texttt{"}, \texttt{"}latent\_image\texttt{"}], [\texttt{"}Joy\_caption\_0\texttt{"}, \texttt{"}STRING\texttt{"}, \texttt{"}ReplaceString\_0\texttt{"}, \texttt{"}ReplaceWith\texttt{"}], [\texttt{"}LayerUtility: ImageRemoveAlpha\_0\texttt{"}, \texttt{"}RGB\_image\texttt{"}, \texttt{"}ImageResizeKJ\_0\texttt{"}, \texttt{"}image\texttt{"}], [\texttt{"}LayerUtility: ImageRemoveAlpha\_0\texttt{"}, \texttt{"}RGB\_image\texttt{"}, \texttt{"}Joy\_caption\_0\texttt{"}, \texttt{"}image\texttt{"}], [\texttt{"}VAEDecode\_0\texttt{"}, \texttt{"}IMAGE\texttt{"}, \texttt{"}ImageCrop\_0\texttt{"}, \texttt{"}image\texttt{"}], [\texttt{"}easy imageSize\_0\texttt{"}, \texttt{"}width\_int\texttt{"}, \texttt{"}ImageCrop\_0\texttt{"}, \texttt{"}width\texttt{"}], [\texttt{"}easy imageSize\_0\texttt{"}, \texttt{"}height\_int\texttt{"}, \texttt{"}ImageCrop\_0\texttt{"}, \texttt{"}height\texttt{"}], [\texttt{"}easy imageSize\_0\texttt{"}, \texttt{"}width\_int\texttt{"}, \texttt{"}ImageCrop\_0\texttt{"}, \texttt{"}x\texttt{"}], [\texttt{"}ImageCrop\_0\texttt{"}, \texttt{"}IMAGE\texttt{"}, \texttt{"}SaveImage\_0\texttt{"}, \texttt{"}images\texttt{"}], [\texttt{"}LoraLoader\_0\texttt{"}, \texttt{"}MODEL\texttt{"}, \texttt{"}ModelSamplingFlux\_0\texttt{"}, \texttt{"}model\texttt{"}], [\texttt{"}CLIPTextEncodeFlux\_0\texttt{"}, \texttt{"}CONDITIONING\texttt{"}, \texttt{"}InpaintModelConditioning\_0\texttt{"}, \texttt{"}positive\texttt{"}], [\texttt{"}ImageResizeKJ\_0\texttt{"}, \texttt{"}IMAGE\texttt{"}, \texttt{"}ImagePadForOutpaint\_0\texttt{"}, \texttt{"}image\texttt{"}], [\texttt{"}easy imageSize\_0\texttt{"}, \texttt{"}width\_int\texttt{"}, \texttt{"}ImagePadForOutpaint\_0\texttt{"}, \texttt{"}right\texttt{"}], [\texttt{"}ImagePadForOutpaint\_0\texttt{"}, \texttt{"}IMAGE\texttt{"}, \texttt{"}easy imageSize\_1\texttt{"}, \texttt{"}image\texttt{"}], [\texttt{"}ImagePadForOutpaint\_0\texttt{"}, \texttt{"}MASK\texttt{"}, \texttt{"}InpaintModelConditioning\_0\texttt{"}, \texttt{"}mask\texttt{"}], [\texttt{"}easy imageSize\_1\texttt{"}, \texttt{"}width\_int\texttt{"}, \texttt{"}LayerUtility: ColorImage\_0\texttt{"}, \texttt{"}width\texttt{"}], [\texttt{"}easy imageSize\_1\texttt{"}, \texttt{"}height\_int\texttt{"}, \texttt{"}LayerUtility: ColorImage\_0\texttt{"}, \texttt{"}height\texttt{"}], [\texttt{"}LayerUtility: ColorImage\_0\texttt{"}, \texttt{"}image\texttt{"}, \texttt{"}ImageComposite+\_0\texttt{"}, \texttt{"}destination\texttt{"}], [\texttt{"}ImageResizeKJ\_0\texttt{"}, \texttt{"}IMAGE\texttt{"}, \texttt{"}ImageComposite+\_0\texttt{"}, \texttt{"}source\texttt{"}], [\texttt{"}ImageComposite+\_0\texttt{"}, \texttt{"}IMAGE\texttt{"}, \texttt{"}InpaintModelConditioning\_0\texttt{"}, \texttt{"}pixels\texttt{"}], [\texttt{"}LoadImage\_0\texttt{"}, \texttt{"}IMAGE\texttt{"}, \texttt{"}LayerUtility: ImageRemoveAlpha\_0\texttt{"}, \texttt{"}RGBA\_image\texttt{"}]]\\
        \hline
        \textbf{RefineAgent}\\
        \hline
        \small
        \#1 \textbf{Error Detection} - Incorrect Nodes: [ReplaceString]\\
        \small
        \#2 \textbf{Retrieval} - Candidate Nodes: [{\texttt{"}node\_name\texttt{"}: \texttt{"}String Replace\texttt{"}, \texttt{"}input\_names\texttt{"}: [\texttt{"}Text\texttt{"}, \texttt{"}Pattern\texttt{"}, \texttt{"}Replace\_With\texttt{"}, \texttt{"}Mode\texttt{"}], \texttt{"}output\_names\texttt{"}: [\texttt{"}TEXT\texttt{"}]}, {\texttt{"}node\_name\texttt{"}: \texttt{"}LogicUtil\_ReplaceString\texttt{"}, \texttt{"}input\_names\texttt{"}: [\texttt{"}String\texttt{"}, \texttt{"}Regex\texttt{"}, \texttt{"}ReplaceWith\texttt{"}], \texttt{"}output\_names\texttt{"}: [\texttt{"}STRING\texttt{"}]}, {\texttt{"}node\_name\texttt{"}: \texttt{"}replace\_string\texttt{"}, \texttt{"}input\_names\texttt{"}: [\texttt{"}input\_string\texttt{"}, \texttt{"}old\_string\texttt{"}, \texttt{"}new\_string\texttt{"}], \texttt{"}output\_names\texttt{"}: [\texttt{"}string\texttt{"}]}, {\texttt{"}node\_name\texttt{"}: \texttt{"}String Replace (mtb)\texttt{"}, \texttt{"}input\_names\texttt{"}: [\texttt{"}string\texttt{"}, \texttt{"}old\texttt{"}, \texttt{"}new\texttt{"}], \texttt{"}output\_names\texttt{"}: [\texttt{"}STRING\texttt{"}]}, {\texttt{"}node\_name\texttt{"}: \texttt{"}replace String \_O\texttt{"}, \texttt{"}input\_names\texttt{"}: [\texttt{"}string\texttt{"}, \texttt{"}old\texttt{"}, \texttt{"}new\texttt{"}], \texttt{"}output\_names\texttt{"}: [\texttt{"}STR\texttt{"}]}]\\
        \small
        \#2 \textbf{LLM Judgment} - Replaces: [\texttt{"}ReplaceString\texttt{"}: \texttt{"}LogicUtil\_ReplaceString\texttt{"}]\\
        
        \hline
    \end{tabular}}
    \caption{\textbf{The process of the ComfyUI workflow generation in ComfyGPT.}}
    \label{tab:over}
\end{table*}

\begin{table*}[t]
    \centering
    \renewcommand{\arraystretch}{1.2} 
    \scalebox{1}{\begin{tabular}{|p{14cm}|}
        \hline
        \textbf{RefineAgent} \\ 
        \hline
        \small
        \#\textbf{Prompt} - I would like you to act as an expert in ComfyUI platform. I will provide a example, including a description about ComfyUI workflow and a logical diagram in json format represents the comfyui workflow. The logical diagram is a links list [link\_1, link\_2, link\_3, ... , link\_n],  each link is consist of [output\_node\_name, output\_name, input\_node\_name, input\_name], represents a line between output node and input node. Example: Description: \{\{\textit{Description}\}\}. Logical Diagram: \{\{\textit{Diagram}\}\}. Now, This logical diagram has one error node name. Error Name: \{\{\textit{Name}\}\}. I will give you some candidate nodes. Please combine thse above information to select the most suitable candidate node. Candidate nodes: \{\{\textit{Nodes}\}\}. You just need to return you choose node name. Please return result in pure JSON format, including: 
      
        \small
        \texttt{```}json\{\texttt{"}candidate\_node\_name\texttt{"}: ...\}\texttt{```}\\
        \hline
    \end{tabular}}
    \caption{\textbf{The prompt design of RefineAgent.} In the prompt, we define some injectable slots, such as \{\{\textit{Description}\}\} and \{\{\textit{Diagram}\}\}, which are dynamically replaced by specific content during execution.}
    \label{tab:refineagent}
\end{table*}

      

\begin{table*}[t]
    \centering
    \renewcommand{\arraystretch}{1.2} 
    \scalebox{1}{
    \begin{tabular}{|p{14cm}|}
        \hline
        \textbf{Few-Shot Learning} \\ 
        \hline
        \small
        \#\textbf{Prompt} - I would like you to act as an expert in ComfyUI platform. I will provide some examples, including a description about ComfyUI workflow and a logical diagram in json format represents the comfyui workflow. The logical diagram is a edges list [edge\_1, edge\_2, edge\_3, ... , edge\_n],  each edge is consist of [output\_node,output\_name,input\_node,input\_name], represents a line between output node and input node. Examples: \{\{\textit{Examples}\}\}. Now, I want you to understand these example and create a new diagram based on a new description. Description: \{\{\textit{Description}\}\} Notes: 1. You only should return the diagram in pure json format without including any other information or code. Response example:\\
        
        \texttt{```}json\{\texttt{"}diagram\texttt{"}: .... \}\texttt{```}\\

        \hline
    \end{tabular}
    }
    \caption{\textbf{The prompt design of the few-shot learning.} In the prompt, we define some injectable slots, such as \{\{\textit{Examples}\}\} and \{\{\textit{Description}\}\}, which are dynamically replaced by specific content during execution.}
    \label{tab:one-shot}
\end{table*}

\begin{table*}[t]
    \centering
    \renewcommand{\arraystretch}{1.2} 
    \scalebox{1}{\begin{tabular}{|p{14cm}|}
        \hline
        \textbf{PIA Evaluation} \\ 
        \hline
        \small
        \#\textbf{Prompt} - I would like you to act as an expert in ComfyUI platform. I will provide a example, including a description about ComfyUI workflow and a ComfyUI nodes list.  Example: Description: \{\{\textit{Description}\}\} ComfyUI Nodes List: \{\{\textit{Nodes}\}\}. Now, I want you to determine whether these ComfyUI nodes can complete the description. Notes: 1. You only need to answer yes or no,  without including any other information or code.\\

        \hline
    \end{tabular}}
    \caption{\textbf{The prompt design of the evaluation of Pass Instruct Alignment(PIA).} In the prompt, we define some injectable slots, such as \{\{\textit{Description}\}\} and \{\{\textit{Nodes}\}\}, which are dynamically replaced by specific content during execution.}
    \label{tab:confirm}
\end{table*}

\begin{figure*}[t]
  \centering
  \includegraphics[width=0.95\linewidth]{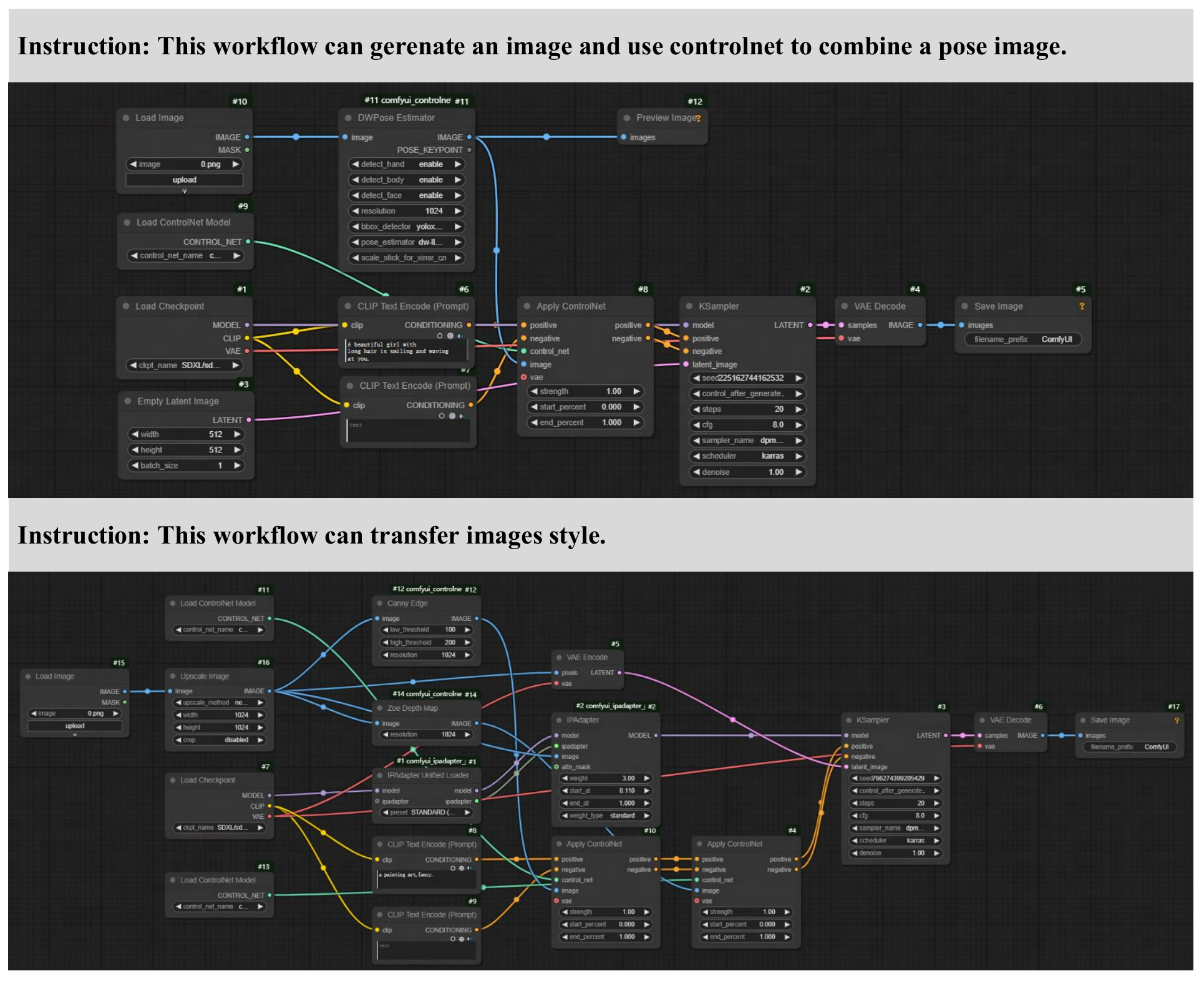}
  \caption{\textbf{The specific ComfyUI workflow is illustrated in Fig.~5 (in the main paper) (b) and (c).}}
  \label{fig:example_1}
\end{figure*}

\begin{figure*}[t]
  \centering
  \includegraphics[width=0.95\linewidth]{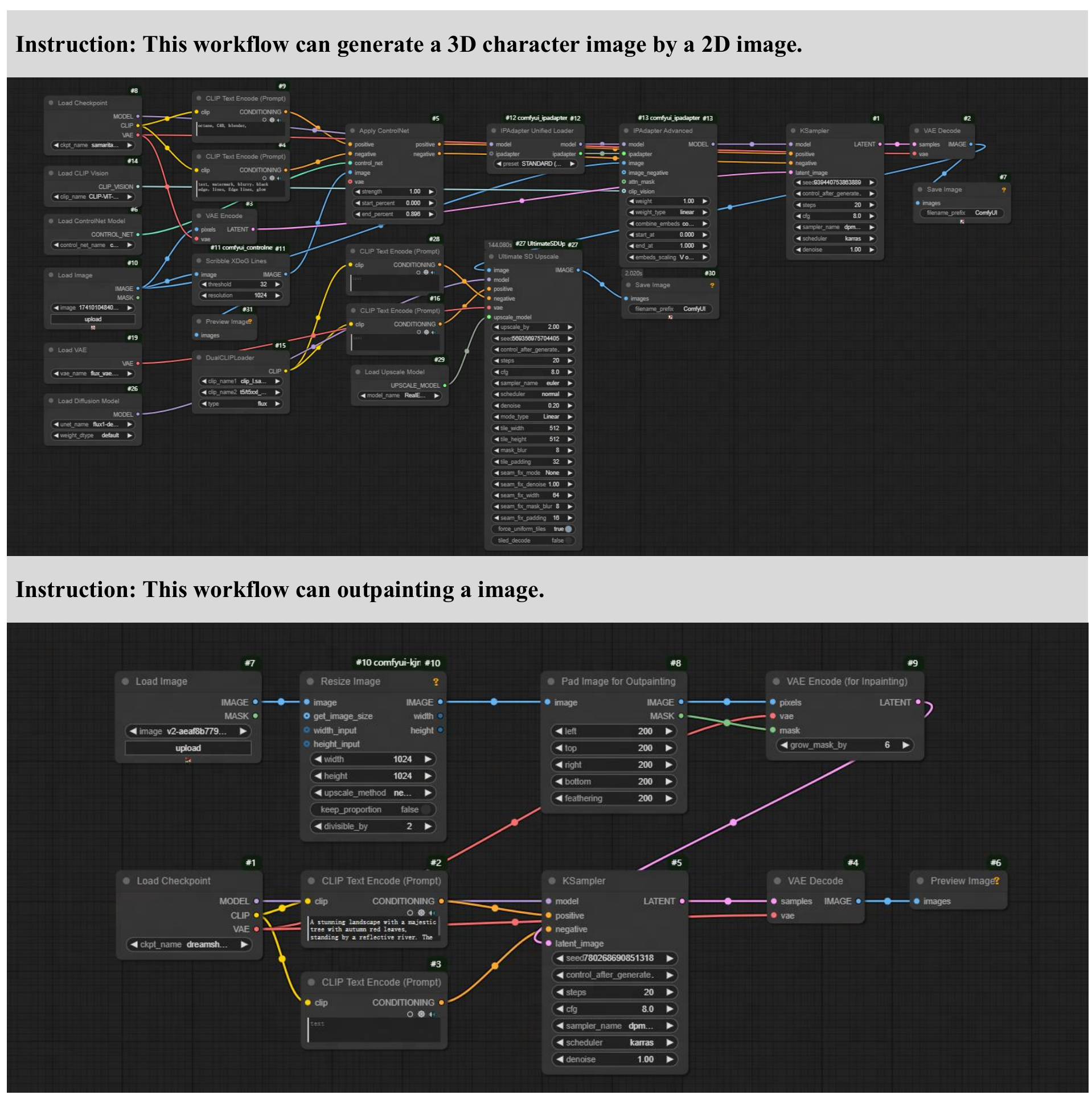}
  \caption{\textbf{The specific ComfyUI workflow is illustrated in Fig.~5(in the main paper) (d) and (e).}}
  \label{fig:example_2}
\end{figure*}


\end{document}